\documentclass{jfm}
\usepackage{graphicx}
\usepackage{epstopdf, epsfig}
\usepackage{dcolumn}
\usepackage{bm}
\usepackage{amsmath,rotating}
\usepackage{amssymb}
\usepackage{dashrule}
\usepackage{setspace}
\usepackage{enumerate}
\usepackage{subcaption}
\usepackage{amsfonts,amsmath}
\usepackage{color}
\usepackage{xfrac}
\usepackage{tikz,url}
\usepackage{float}
\usepackage{upgreek}
\usepackage{enumitem}
\usetikzlibrary{shapes.misc}
\usepackage{diagbox}
\usepackage{titlesec}
\usepackage{mathrsfs}
\usepackage{array}
\usepackage[normalem]{ulem}
\numberwithin{equation}{section}

\tikzset{cross/.style={cross out, draw=black, minimum size=2*(#1-\pgflinewidth), inner sep=0pt, outer sep=0pt},cross/.default={1pt}} 

\usetikzlibrary{shapes}

\DeclareMathSizes{10}{10}{7}{5}

\shorttitle{Inertial Particle Clustering in a Porous Cell}
\shortauthor{S. Apte, T. Oujia, K. Matsuda, B. Kadoch, X. He, and K. Schneider}

\title{\textcolor{black}{Clustering of Inertial Particles in Turbulent Flow Through a Porous Unit Cell}}
\author{Sourabh V. Apte\aff{1},
\corresp{\email{Sourabh.Apte@oregonstate.edu }},
Thibault Oujia\aff{2},
Keigo Matsuda\aff{3},
Benjamin Kadoch\aff{4},
Xiaoliang He\aff{5},
 \and Kai Schneider\aff{2}
 }

\affiliation{\aff{1}School of Mechanical, Industrial and Manufacturing Engineering, Oregon State University, Corvallis, OR 97330, USA
\aff{2} Aix--Marseille Universit\'e, I2M--CNRS, Marseille, France
\aff{3} Research Institute for Value-Added-Information Generation (VAiG), Japan Agency for Marine-Earth Science and Technology (JAMSTEC), Yokohama 236-0001, Japan
\aff{4} Aix--Marseille Universit\'e, IUSTI--CNRS, Marseille, France
\aff{5} Pacific Northwest National Laboratory, Richland, WA 99352, USA
}

\begin{document}

\newcommand{\linetriangle}{\raisebox{0pt}{\tikz{\node[draw,scale=0.4,regular polygon, regular polygon sides=3,fill=none,rotate=0](){};\draw[-,black,dashed,line width = 1.0pt](-5mm,0.0mm) -- (5mm,0.0mm)}}}

\newcommand{\linecircle}{\raisebox{0pt}{\tikz{\node[draw,scale=0.65,circle,fill=none,rotate=180](){};\draw[-,black,solid,line width = 1.0pt](-5mm,0.0mm) -- (5mm,0.0mm)}}}

\newcommand{\redline}{\raisebox{2pt}{\tikz{\draw[-,red,dashed,line width = 1.5pt](0,0) -- (8mm,0);}}}

\newcommand{\dotline}{\raisebox{2pt}{\tikz{\draw[-,black,dotted,line width = 1.5pt](0,0) -- (8mm,0);}}}

\newcommand{\dashdotline}{\raisebox{2pt}{\tikz{\draw [-,black,dashdotted,line width = 1.25pt] (0,0) -- (10mm,0);}}}

\newcommand{\linex}{\raisebox{0pt}{\tikz{\node[draw,scale=5,cross,fill=none,rotate=180](){};\draw[-,black,solid,line width = 1.0pt](-5mm,0.0mm) -- (5mm,0.0mm)}}}

\newcommand{\mytriangle}{\raisebox{0.5pt}{\tikz{\node[draw,scale=0.4,regular polygon, regular polygon sides=3,fill=none,rotate=0](){};}}}

\newcommand{\mycircle}{\raisebox{0.5pt}{\tikz{\node[draw,scale=0.5,circle,fill=none](){};}}}

\newcommand{\mydiamond}{\raisebox{0pt}{\tikz{\node[draw,scale=0.4,diamond,fill=none](){};}}}

\newcommand{\blue}{\textcolor{black}}
\newcommand{\red}{\textcolor{red}}

%
\newcommand{\twoform}[1]{\mathbfsf{#1}}

\newcommand*\Sourabh[1]{\textcolor{red}{#1}}
\newcommand*\Thibault[1]{\textcolor{cyan}{#1}}
\newcommand*\Benjamin[1]{\textcolor{green}{#1}}
\newcommand*\Keigo[1]{\textcolor{purple}{#1}}
\newcommand*\Bryan[1]{\textcolor{brown}{#1}}
\newcommand*\Kai[1]{\textcolor{blue}{#1}}

\newenvironment{enumerate*}%
  {\begin{enumerate}%
    \setlength{\itemsep}{0pt}%
    \setlength{\parskip}{0pt}}%
  {\end{enumerate}}

\setcounter{secnumdepth}{4}

\titleformat{\paragraph}
{\normalfont\normalsize\bfseries}{\theparagraph}{1em}{}
\titlespacing*{\paragraph}
{0pt}{3.25ex plus 1ex minus .2ex}{1.5ex plus .2ex}

\maketitle

\begin{abstract}
Direct numerical simulation is used to investigate effects of turbulent flow in the confined geometry of a face-centered cubic porous unit cell on the transport, clustering, and deposition of fine particles at different Stokes numbers ($St = 0.01, 0.1, 0.5, 1, 2$) and at a pore Reynolds number of 500. Particles are advanced using one-way coupling and collision of particles with pore walls is modeled as perfectly elastic with specular reflection. Tools for studying inertial particle dynamics and clustering developed for homogeneous flows are adapted to take into account the embedded, curved geometry of the pore walls. The pattern and dynamics of clustering are investigated using the volume change of Voronoi tesselation in time to analyze the divergence and convergence of the particles. Similar to the case of homogeneous, isotropic turbulence, the cluster formation is present at large volumes, while cluster destruction is prominent at small volumes and these effects are amplified with Stokes number. However, unlike homogeneous, isotropic turbulence, formation of large number of very small volumes was observed at all Stokes numbers and is attributed to the collision of particles with the pore wall. Multiscale wavelet analysis of the particle number density showed peak of clustering shifts towards larger scales with increase in Stokes number. Scale-dependent skewness and flatness quantify the intermittent void and cluster distribution, with cluster formation observed at small scales for all Stokes numbers, and void regions at large scales for large Stokes numbers.
\end{abstract}


\maketitle

\medskip

%

\singlespacing

\section{Introduction}\label{intro}
Several applications including catalysis in the chemical synthesis and process industries~\citep{Dixon2001aa,aris1999elementary}, high temperature nuclear reactor cooling~\citep{shams2013quasi},  hyporheic exchange of pollutants at the sediment-water interface~\citep{hester2017importance}, sand-migration in oil/gas wells, involve unsteady transitional and turbulent flows through confined spaces and porous structures.
In these examples, the contribution of the inertial terms in the fluid flow equations can dramatically change the topology of the flow field leading to formation of jets, vortices, dead zones, etc. within the pores. Such flow features can substantially alter the dispersion characteristics of pollutants and play critical roles in the transport of reactants and products to and from active reaction sites. 

Dynamics of small inertial particles, such as sand particles, through porous media is of importance in oil and gas production~\citep{carlson1992sand}. Sand production is considered one of the most important issues facing hydrocarbon wells as it can erode hardware, plug tubulars, cease flow, create downhole cavities, and needs separation and disposal at surface. Variations in the reservoir pressure and completion permeability leads to high velocity in the perforation tunnels, significant inertial and turbulence effects and kinetic energy losses in the completion region~\citep{cook2004discrete}, and onset of sanding triggered by rock failure~\citep{yi2005effect}. Minimizing pressure drop means that the gravel bed porosity should be as large as possible. However, to act as an effective filter for sand grains, the gravel also has to be small enough to restrain formation sand~\citep{saucier1974considerations,mahmud2019sand}. Particle rentention, bridge formation, jamming, and deposition has been studied experimentally in simplified configurations such as microchannels~\citep{valdes2006particle,dai2010blockage,agbangla2012experimental} and packed beds~\citep{ramachandran1999plugging,pandya1998existence} to show that rate of plugging by bridging has a nonlinear dependence on particle concentration. This bridging effect depends on flow velocity, particle diameter-to-pore size ratio, and flow geometry. 

The goal of the present work is to investigate and understand the clustering dynamics of inertial particles in a turbulent flow through confined geometries representative of porous medium. Specifically, how does the interaction of particles with the solid walls affect the clustering and deposition and how this changes with particle Stokes number is of critical importance. In addition, with high Reynolds number flows through porous media, the pore-scale flow structure can change significantly owing to the inertial and unsteady flow features. Previous work~\citep{he2019characteristics} explored, using direct numerical simulation (DNS), how this change of flow structure impacts the turbulence (i.e. the turbulent kinetic energy distribution and transport mechanisms) in a face-centered cubic (FCC) lattice with very low porosity. The flow geometry gives rise to rapid acceleration and deceleration of the mean flow in different regions with presence of three-dimensional (3D) helical motions, weak wake-like structures behind the bead spheres, stagnation and jet-impingement-like flows together with merging and spreading jets in the main pore space. The jet-impingement and weak wake-like flow structures give rise to regions with negative production of total turbulent kinetic energy, a feature observed in jet-impingement like configurations. Unlike flows in complex shaped ducts, the turbulence intensity levels in the cross-stream directions were found to be larger than those in the streamwise direction. Furthermore, due to the compact nature and confined geometry of the FCC packing, the turbulent integral length scales were estimated to be less than 10\% of the bead diameter even for the lowest, transitional Reynolds number, indicating the absence of macroscale turbulence structures for this configuration. This finding suggests that even for the highly anisotropic flow within the pore, the upscaled flow statistics are captured well by the representative volumes defined by the unit cell. In the present work, the previous analysis of turbulence in porous geometry is extended to investigate dynamics and transport of inertial particles and quantify the effect of geometric confinement on particle clustering.

Clustering of inertial particles in turbulent flows has been well studied in homogeneous isotropic turbulence and wall-bounded channel flows~\citep{maxey1987gravitational,eaton1994preferential,toschi2009lagrangian,monchaux2012analyzing}. The divergence of particle velocity, which differs from the divergence-free fluid velocity in an incompressible flow, plays a crucial role in clustering of inertial particles. The divergence of particle velocity has been shown to be proportional to the second invariant of flow velocity gradient tensor for sufficiently small Stokes numbers, which is defined as the ratio of the particle relaxation time, $\tau_p$, to the Kolmogorov time, $\tau_{\eta}$. Particles tend to concentrate in low vorticity and high strain regions in turbulence resulting in the preferential concentration. Divergence of particle velocity has been used to quantify clustering mechanisms~\citep{esmaily2016analysis,bec2007heavy}. An inherent difficulty for determining the divergence of the particle velocity is its discrete nature, i.e. it is only defined at particle positions. Recently,~\citet{oujia2020jfm} computed the particle velocity divergence from the position and velocity of a large number of particles, using a Voronoi tessellation technique. They proposed a model for quantifying the divergence using tessellation of the particle positions. The corresponding time change of the volume is shown to yield a measure of the particle velocity divergence.

Pair correlation function (PCF) has also been widely used to analyze clustering as it is directly related to the particle collision rates~\citep{sundaram1997collision,wang2000statistical}. The PCF typically shows a power-law behavior at sub-Kolmogorov scales and the slope is dependent on the Stokes number. Scale similarity of particle distribution has been explained by the sweep-stick mechanism proposed by~\cite{goto2006self} in which particles are swept by large-scale 
flow motion while sticking to stagnation points of Lagrangian fluid acceleration~\citep{coleman2009unified}. Probability distribution function (PDF) of particle mass density and coarse graining techniques have been used to investigate scale dependence of particle distribution~\citep{bec2007heavy}. ~\citet{bassenne2017extraction} proposed a wavelet-based method to extract coherent clusters of inertial particles in fully developed turbulence. Wavelets decompose turbulent flow,  scalar and vector valued fields, in scale, position, and possibly direction, complementary to Fourier techniques which yield insight into wave number contributions of turbulent flow fields. The wavelet representation can be used to analyze spatial intermittency and quantify spatial fluctuations at different scales. This is not easily possible with Fourier transform, owing to the global character of the basis functions. For a review on wavelets and turbulence refer to \cite{farge1992,schneidervasilyev2010} and more particularly on wavelet-based statistics to~\cite{farge2015}. Recently,~\citet{matsuda2020scale} obtained scale-dependent statistics of the particle distribution and insights into the multiscale structure of clusters and voids in particle-laden, homogeneous isotropic turbulent flow using orthogonal wavelet decomposition of the Eulerian particle density field at high Reynolds numbers.

Although particle clustering has been observed and studied in several particle-laden turbulent flows, dynamics and clustering of small, inertial particles in complex and confined configurations of densely packed porous beads has not been explored. Whether the commonly observed heavy particle clustering mechanisms at unit particle Stokes numbers also appear in a confined geometry, and how collisional interactions of these particles with the bead surfaces affect such clustering has not been studied. To address these issues, DNS is used to investigate effect of turbulent flow in the confined geometry of a face-centered cubic porous unit cell on the transport of fine, inertial particles at different Stokes numbers ($St = 0.01, 0.1, 0.5, 1, 2$) and at a pore Reynolds number of 500. Particles are advanced using one-way coupling and collision of particles with pore walls is modeled as perfectly elastic specular reflection. Detailed analysis of clustering and void formation statistics is then conducted based on (i) Voronoi tessellation, and (ii) wavelet analysis of particle number density field in the porous geometry. The clustering statistics are compared and contrasted against those in a homogeneous, isotropic turbulence flow~\citep{oujia2020jfm,matsuda2020scale}. 
Specifically, the impact of geometric confinement on particle clustering and void formation at different Stokes numbers is evaluated.


The rest of the paper is arranged in the following way. Section~\ref{setup} provides the details of the porous geometry, simulation parameters, the numerical approach, as well as Lagrangian tracking and motion of inertial particles. Mean velocity field, TKE distributions, integral scales are described in Section~\ref{results}. Analysis of particle clusters and voids is then conducted using Voronoi tessellation and wavelet-based multiscale, scale dependent statistics of particle number density. Finally, summary and conclusions are given in Section~\ref{summ}.

\begin{figure}
    	\centering
    	\begin{subfigure}[b]{0.4\textwidth}
    	    \centering
            \includegraphics[width=\textwidth]{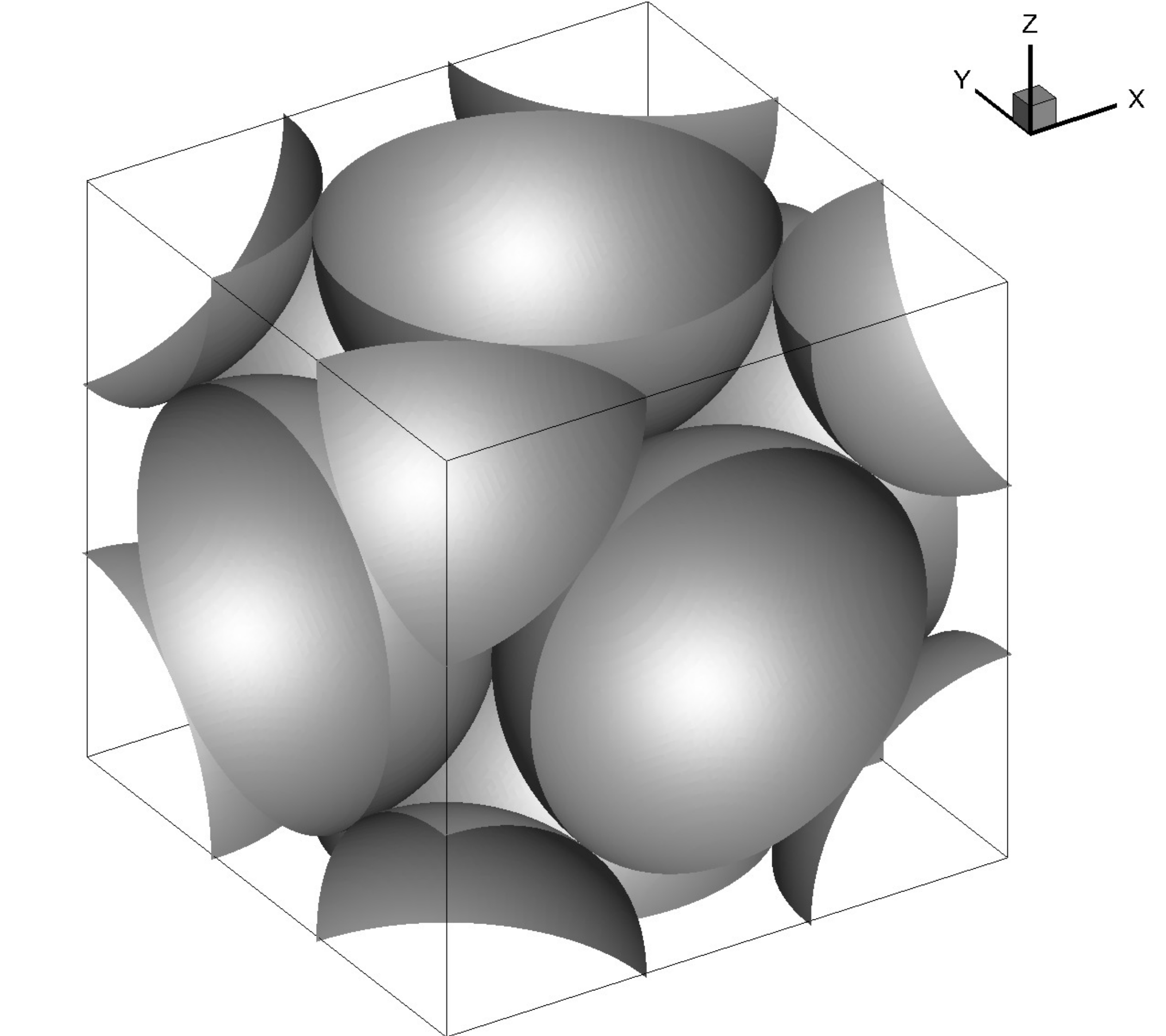}
            \caption{~}
        	\label{fig_geoa}
        \end{subfigure}
        \begin{subfigure}[b]{0.4\textwidth}
           \centering
           \includegraphics[width=\textwidth]{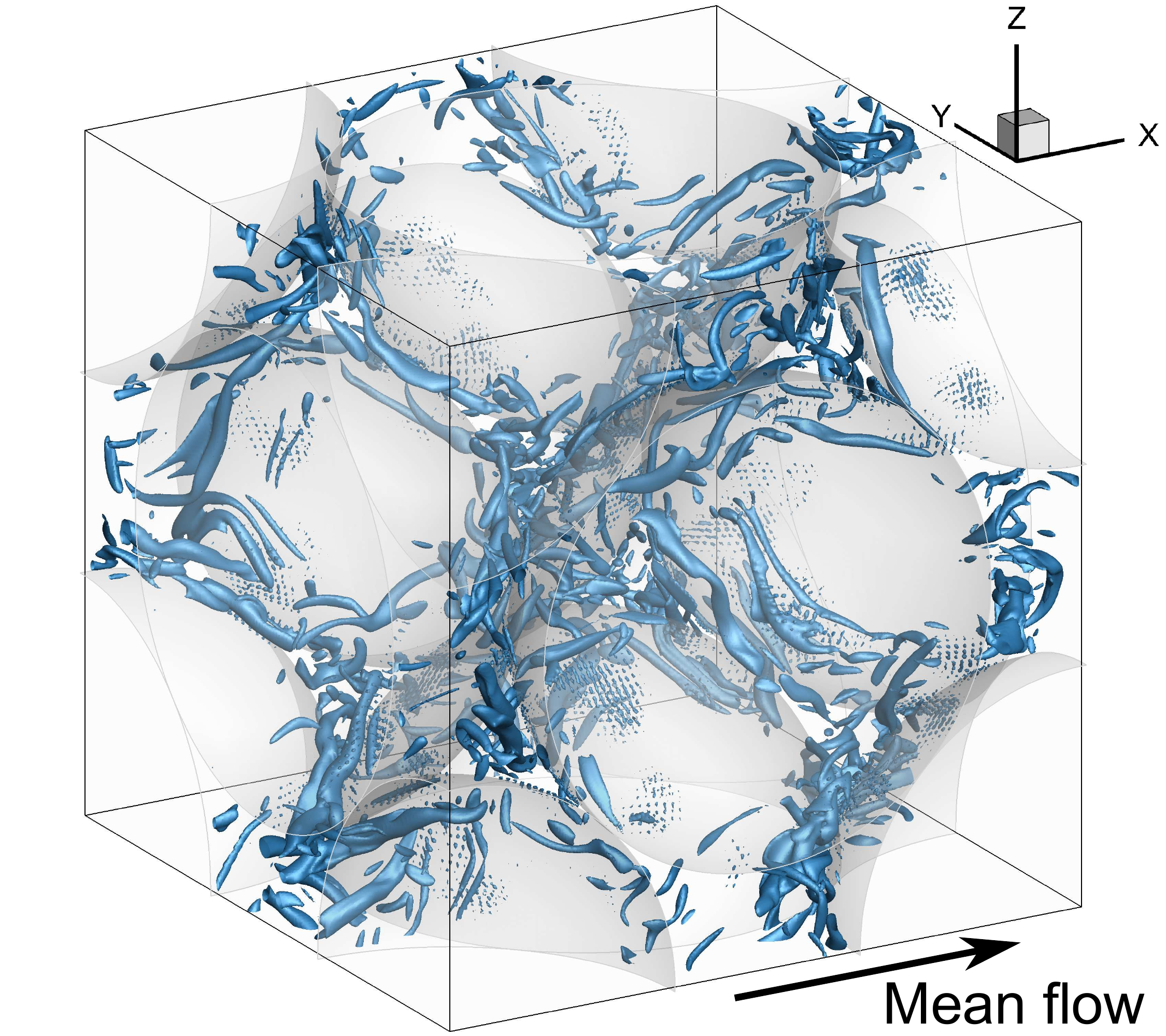}
           \caption{~}
            \label{fig_geob}
        \end{subfigure}
	\caption{(a) Schematic of a face-centered porous unit cubic cell showing the bead arrangement and surface geometry, (b) the isosurface of swirling strength at $\lambda = 0.25 \lambda_{max}$ for $Re_H = 500$.}
	\label{fig_geo}
\end{figure}

\section{Simulation setup}\label{setup}
Several different definitions of Reynolds numbers have been used in porous media literature~\citep{he2019characteristics,wood2020modeling} based on different length scales such as the particle diameter ($D_B$), the hydraulic diameter ($D_H$), or the permeability. In this work, Reynolds number ($Re_H$) based on $D_H$ is used. 
The hydraulic diameter is related to the particle diameter as,
\begin{equation}
\label{eq:dh}
D_H = {\frac{2}{3}} {\frac{\phi}{1-\phi}} D_B.
\end{equation}
Then, $Re_H$ is defined as (dropping the factor 2/3),
\begin{equation}
		Re_H = \frac{\overline{\langle u_x \rangle ^f} D_B}{\nu_f} 
		\label{eq:rep}
\end{equation}
where $\nu_f$ is the kinematic viscosity of fluid, $\phi$ is the porosity of the medium defined as the ratio of the void 
volume (which corresponds to the fluid volume) to the total volume, $\overline{\langle u_x \rangle ^f}$ is the time-averaged interstitial (intrinsic average) velocity of flow in porous media, and  $u_x$ is the instantaneous velocity component in $x$-(streamwise) direction. The spatial averaging operation is denoted by $\langle~\cdot~ \rangle$, the superscript or subscript $f$ indicates the fluid phase, and $\overline{~\cdot~}$ is the time averaging operator.
For clarity, note that the intrinsic average velocity is defined by,
\begin{equation}
\label{eq:intrinsic}
\left.\langle u_x \rangle^f\right|_t = \frac{1}{V_f}\int_{{\bf x}\in \Omega_f} u_x({\bf x},t)\, dV({\bf x}),
\end{equation}
where $V_f$ is the volume of the fluid phase within the unit cell and $\Omega_f$ denotes the fluid domain. 
The overbar represents traditional time averaging of a quantity $q$ defined as,
\begin{equation}
    \overline{q}(t)=\frac{1}{T}\hspace{-8mm}\int\limits_{~~~~~~t^*=t-T/2}^{~~~~~~t^*=t+T/2} q(t^*)dt^*.
\end{equation}

	\subsection{Porous geometry}\label{geo}
	 A porous face-centered cubic (FCC) unit cell  (Fig.~\ref{fig_geo}) is used based on our prior work on turbulence in porous media~\citep{he2018angular,he2019characteristics}. It has a half sphere entering at each face of the cube, and a half quarter sphere at each corner. The face-centered cubic arrangement creates the lowest possible porosity ($\phi$) 
	 to be 0.26 for the structured packings. Due to this extreme compactness, the flow through the unit cell experiences rapid expansion and contraction. A pressure gradient is imposed to drive the flow through the bed and a triply periodic boundary condition is applied for the unit cell. Majority of the flow enters the cubic cell through the upstream open corners, converges into the center pore resulting in strong accelerations and decelerations, and then leaving the unit cell along downstream corners. In this work, the flow at one pore Reynolds numbers, $Re_H = 500$ is simulated using direct numerical simulation, resulting in a turbulent flow within the pore. Emphasis is placed on dynamics of inertial particles at different Stokes numbers (0.01, 0.1, 0.5, 1, 2), the definition is given below.
%
	\subsection{Numerical Approach and Grid Convergence}\label{grid}
	The numerical approach is based on a fictitious  domain  method  to  handle  arbitrarily  shaped  immersed  objects
without  requiring  the  need  for  body-fitted  grids~(Ref.~\cite{apte2009frs}). 
Uniform Cartesian grids are used in the entire simulation domain, including both fluid and solid phases. An additional body force is imposed on the solid part to enforce the rigidity constraint and satisfy the no-slip boundary condition. The absence of highly skewed unstructured mesh at the bead surface has been shown to accelerate the convergence and lower the uncertainty  (Ref.~\cite{justin2013rela}). 
The following governing equations are solved over the entire domain, including the region within the solid bed, and a rigidity constraint force, $\bf f$, is applied that is non-zero only in the solid region.

The governing equations read as:
	\begin{align}
		\nabla\cdot{\bf u} &= 0, \label{eq:NSa} \\
		\rho_f \bigg[\frac{\partial {\bf u}}{\partial t} + \left({\bf u}\cdot \nabla\right) {\bf u}	\bigg] &= 
		-\nabla p + \mu_f \nabla^2{\bf u} + {\mathbf f} \:, 
	\end{align}
		\label{eq:NSb}
	where $\bf u$ is the velocity vector (with components given by ${\bf u}=(u_x,u_y,u_z)$, $\rho_f$ the fluid density, $\mu_f$ the fluid dynamic viscosity, and $p$ the pressure. 
A fully parallel, structured, collocated grid solver has been developed and thoroughly verified and validated for a range of test cases including flow over a cylinder and sphere for different Reynolds  numbers, flow over touching spheres at different orientations, flow developed by an oscillating cylinder, among others.
The details of the algorithm as well as very detailed verification and validation studies have been published elsewhere (Ref.~\cite{apte2009frs}). In addition, the solver was also used to perform direct one-to-one comparison with a body-fitted solver with known second-order accuracy for steady inertial, unsteady inertial and turbulent flow through porous media (Ref.~\cite{justin2013rela,he2018angular,he2019characteristics}). 

For the present studies, the flow is driven by a pressure drop as a body force in a triply periodic domain.
According to \citet{hill2002trans}, a constant pressure gradient $\nabla P$ in the main direction of the flow ($x$ direction), proportional to the body force $F$, is used to drive the flow,
	\begin{align}
		\partial_x P &=  \frac{18 \: \mu_f \: c \: U_{int}}{ D_B^2} F \:,
		\label{eq:pressgrad}
	\end{align}
where, $\mu_f$ is the dynamic viscosity of the fluid and $c$ the solid volume fraction, defined as $\frac{2}{3} \pi (D_B/L)^3$ ($L$ is the length of the unit cube). The body force $F$ changes with the pore Reynolds number according to the linear fit obtained by~\citet{hill2002trans} and is given as,
	\begin{align}
		F &= 462+9.85\: \left(\frac{1-\phi}{2}Re_H\right)\:~~~ (Re_H \geq 216) \:.
		\label{eq:bodyforce}
	\end{align}
	
A posteriori calculation of body force needed to balance the shear stress on the sphere surfaces for different Reynolds numbers exhibits a good agreement with the above correlation and has been published elsewhere~\citep{he2019characteristics}. A uniform, cubic grid is used with resolution chosen such that the first grid point is at $y^{+}<1$ (i.e., in the viscous sub-layer) to accurately capture the wall-layers, where $y^+=yu_{\tau}/\nu$ indicates the normalized distance from the sphere surface, 	$u_\tau=\sqrt{\tau_\omega/\rho}\approx0.5||\overline{\langle u_x \rangle^f}||$ is the friction velocity, and $\tau_\omega$ is the estimated wall shear stress.
To obtain a more {{direct estimate}} on grid resolution requirements in the present DNS simulations, 3D DNS studies were performed at $Re_H=500$ in a unit cell of face-centered cubic (FCC) spheres with systematic grid refinement study using $48$, $64$, $96$, $112$, $128$ and $144$ grid points per bead diameter $D_B$. A grid converged solution was obtained for first-order (mean flow) as well as second-order statistics (turbulent kinetic energy, TKE) for  $Re_H=500$ at a resolution of $D_B/96$. For $Re_H=500$, the mean flow converged at $D_B/96$, whereas, TKE showed small changes compared to coarser mesh indicating that a grid converged solution can be expected with a resolution of around $D_B/100$. However, in order to obtain a high resolution DNS study and provide sufficient resolution in the bead contact region, a refined grid based on $D_B/\delta = 250$ ($\delta$ is the grid resolution in one direction) was used to resolve the pore-scale flow structures~\citep{he2019characteristics}. 
	

\subsection{Particle tracking algorithm}
\label{sec:part}
A point-particle approach is used to model the motion of small, heavy inertial particles in a Lagrangian frame. In this approach, the particle size is assumed to be much smaller than the Kolmogorov scale, and the forces acting on them are modeled using simple closure models. In the present work, the particle motion is assumed to be governed by a simple, linear drag force. For different Stokes numbers ($St$), the particle motion is given by,
\begin{equation}
\label{eq:lagrangian}
{d {\mathbf x}_p \over dt} = {\mathbf u}_p;~~~ {d {\mathbf u}_p \over dt} = {{\mathbf u}_{f,p} - {\mathbf u}_p \over St\cdot\tau_{\eta}},
\end{equation}
where ${\mathbf x}_p$ and ${\mathbf u}_p$ are particle position and velocity, $\tau_{\eta}$ is Kolmogorov time scale, ${\mathbf u}_{f,p}$ is the 
instantaneous fluid velocity interpolated to the particle location, i.e., ${\bf u}_{f,p}={\bf u}({\bf x}_p)$, and $\tau_p = St~\tau_{\eta}$ is the particle relaxation time.  The particles are advanced using the instantaneous fluid velocity interpolated to the particle location. 
The effect of inhomogeneity in the fluid velocity and the confinement of the bead walls on inertial particle motion can thus be quantified by qualitatively comparing results to inertial particles in homogeneous isotropic turbulence~\citep{matsuda2020scale,oujia2020jfm}.

Particles with five different Stokes numbers, $St=$ 0.01, 0.1, 0.5, 1, and 2 are simulated. One-way coupling, wherein the inertial particles do not affect the fluid flow, is used by assuming that the particle size and concentration is small.
After a stationary turbulent flow is achieved within the porous cell, inertial particles are injected into the fluid domain. One particle is added at the center of each control volume in the fluid region, giving about $N_p \approx 10.4\times 10^6$ 
particles for each Stokes number. The fluid velocity interpolated to the particle location is obtained from tri-linear interpolation, and a fourth-order Runge-Kutta (RK4) scheme is implemented to advance the particle locations in time. Interactions of particles with the bead walls in the unit cell are modeled using Snell's law of specular reflection assuming perfectly elastic collisions. The direction of particle reflection, ${\hat {\mathbf s}}_r$, is determined from the incident direction, ${\hat {\mathbf s}}_i$, and the inward face normal at the spherical bead surface, ${\hat {\mathbf n}}$ as,
\begin{equation}
\label{eq:Snell}
{\hat {\mathbf s}}_r = {\hat {\mathbf s}}_i + 2 |{\hat {\mathbf s}}_i \cdot {\hat{\mathbf n}}| {\hat{\mathbf n}} \, .
\end{equation}
The particle velocity after reflection is also modified according to the above equation. The inward normal to the sphere surface can be easily obtained by knowing the location on the spherical bead surface that the particle crossed and the center of the bead.


\section{Results and discussion}\label{results}
In this section, the mean and turbulent flow structure is described briefly using the the mean and rms (root mean square) velocity fields inside the pore, and the integral length and time scales are estimated based on the Eulerian and Lagrangian two-point auto-correlations, respectively. Next, the probability distribution functions of Voronoi tesselation volumes and the divergence of particle velocity field are evaluated and discussed. In addition, the characteristics of inertial particle clustering in the porous media at various Stokes numbers are discussed using the multiscale, wavelet analyses of the particle number density, wavelet spectra, higher-order statistics of flatness and skewness. The Voronoi tesselation and wavelet analysis results for inertial particle statistics in the FCC unit cell are contrasted with those from isotropic turbulence to identify effects of geometric confinement.

	\subsection{Turbulent flow statistics}\label{flowdes}
As described earlier in the simulation setup, the flow through the triply periodic domain is driven by a body force in the flow direction based on the correlations given by Eq.~(\ref{eq:bodyforce}). After an initial transient, a stationary state is reached and the computation is continued for several flow through times, $T_f = L/U_{int}$ where $L$ is the length of the unit cube. For each case, the flow was first computed for several flow through times to ensure that a stationary state has been reached. This is confirmed by monitoring the total kinetic energy in the domain, which starts out to be a large value and then decreases and remains more or less constant after an initial transient period. This initial transient period was about $100 T_f$ for $Re_H=500$. After a stationary state has been established, the computations were performed for additional $80 T_f$ for each Reynolds number to collect flow statistics which was found to be large enough to obtain converged statistics.

\begin{figure}
     \centering
     \begin{subfigure}[b]{0.47\textwidth}
        \includegraphics[width=\textwidth]{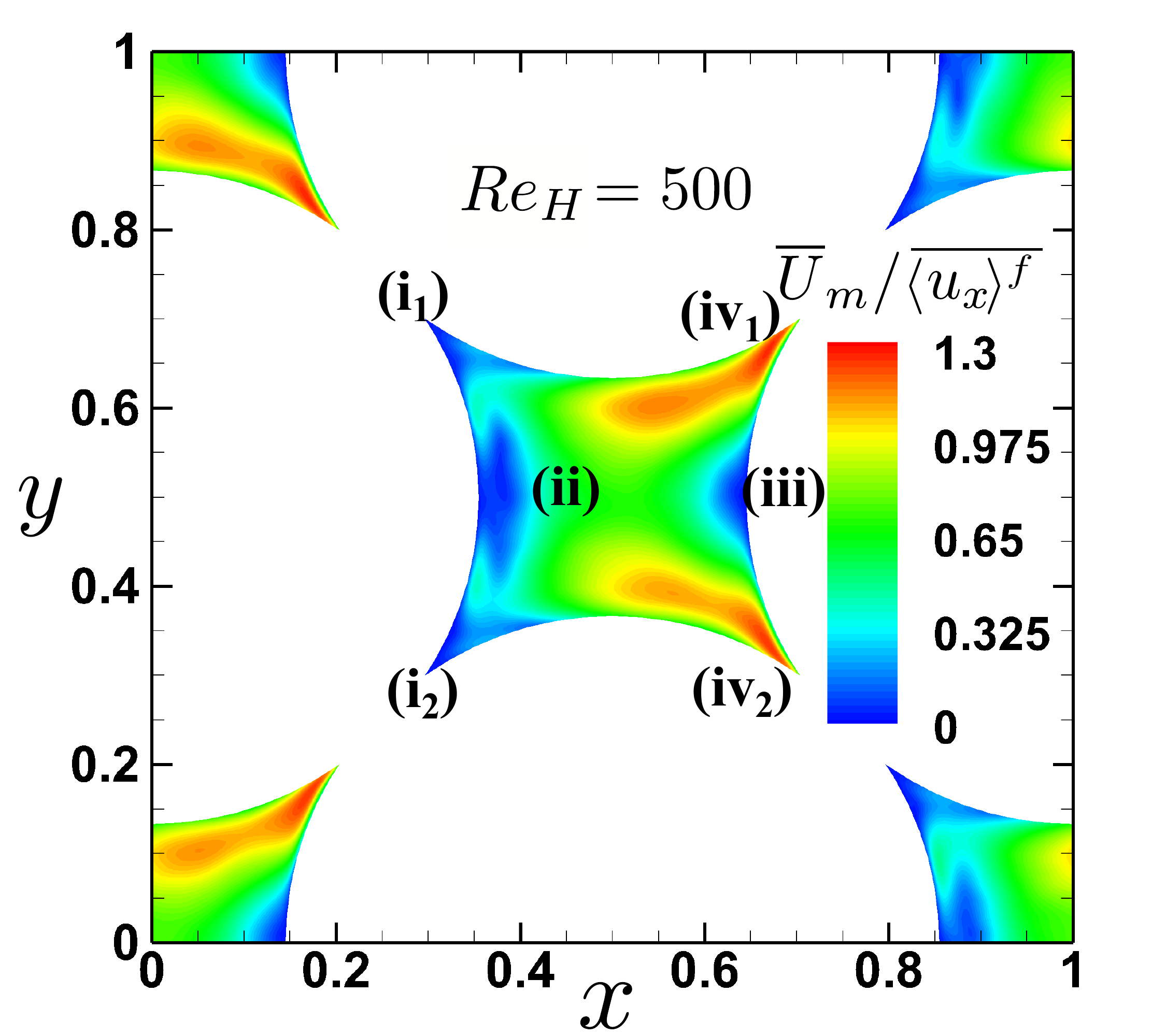}
        \caption{~}
         \label{fig_u500}
    \end{subfigure}
     \begin{subfigure}[b]{0.47\textwidth}
        \includegraphics[width=\textwidth]{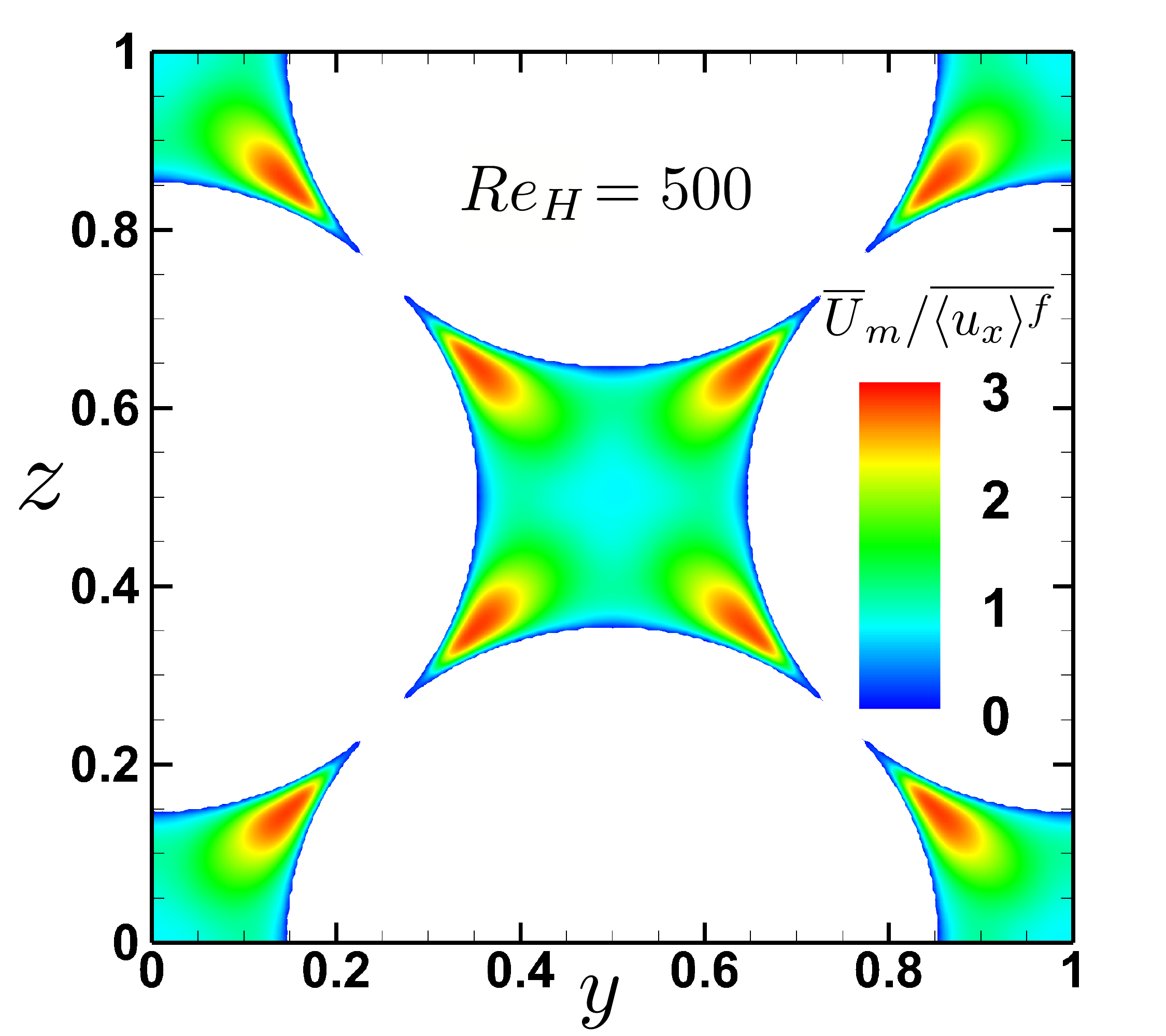}
        \caption{~}
        \label{fig_tke500_xy}
        \end{subfigure}
     \begin{subfigure}[b]{0.47\textwidth}
        \includegraphics[width=\textwidth]{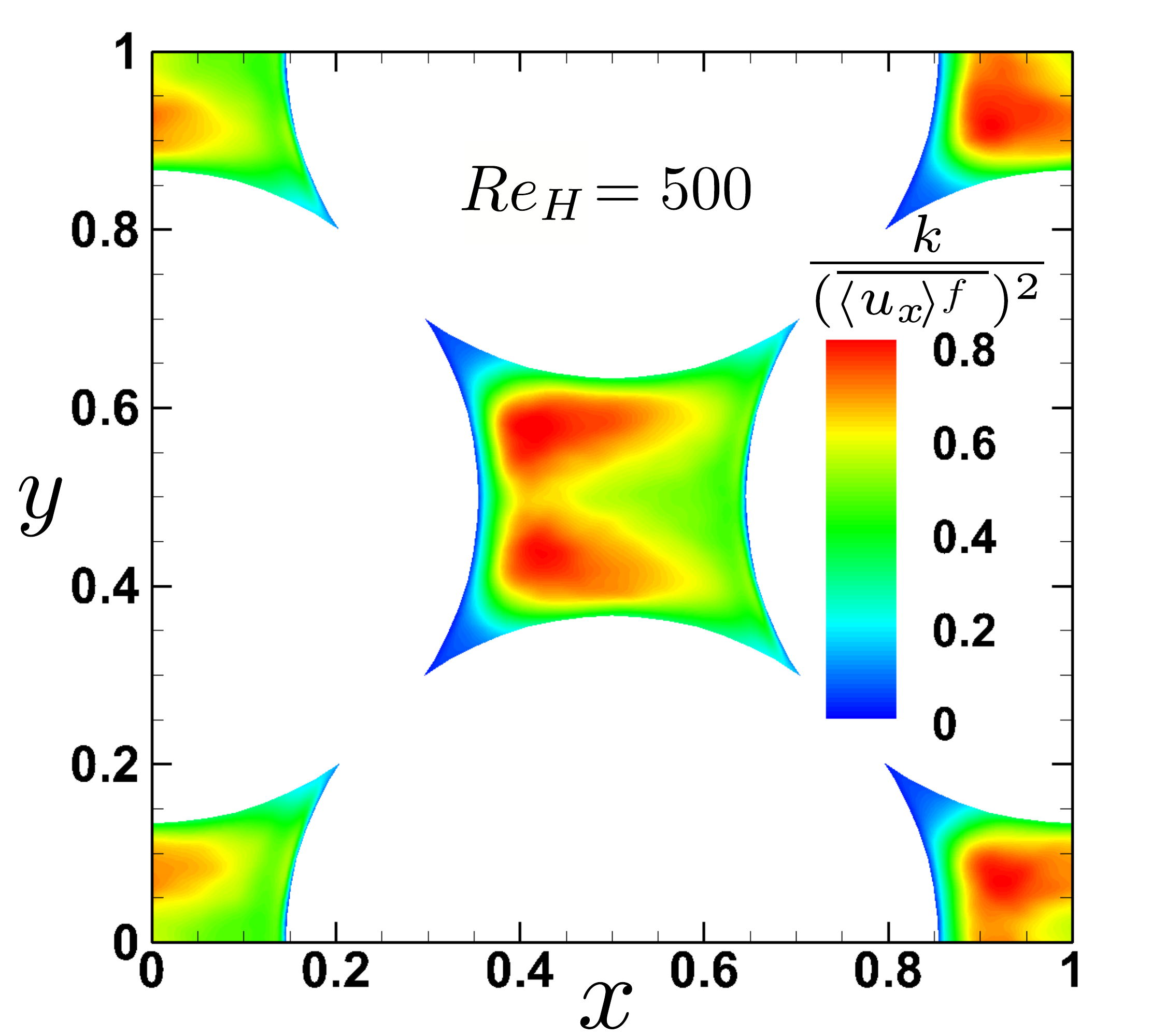}
        \caption{~}
        \label{fig_tke500}
        \end{subfigure}
     \begin{subfigure}[b]{0.47\textwidth}
        \includegraphics[width=\textwidth]{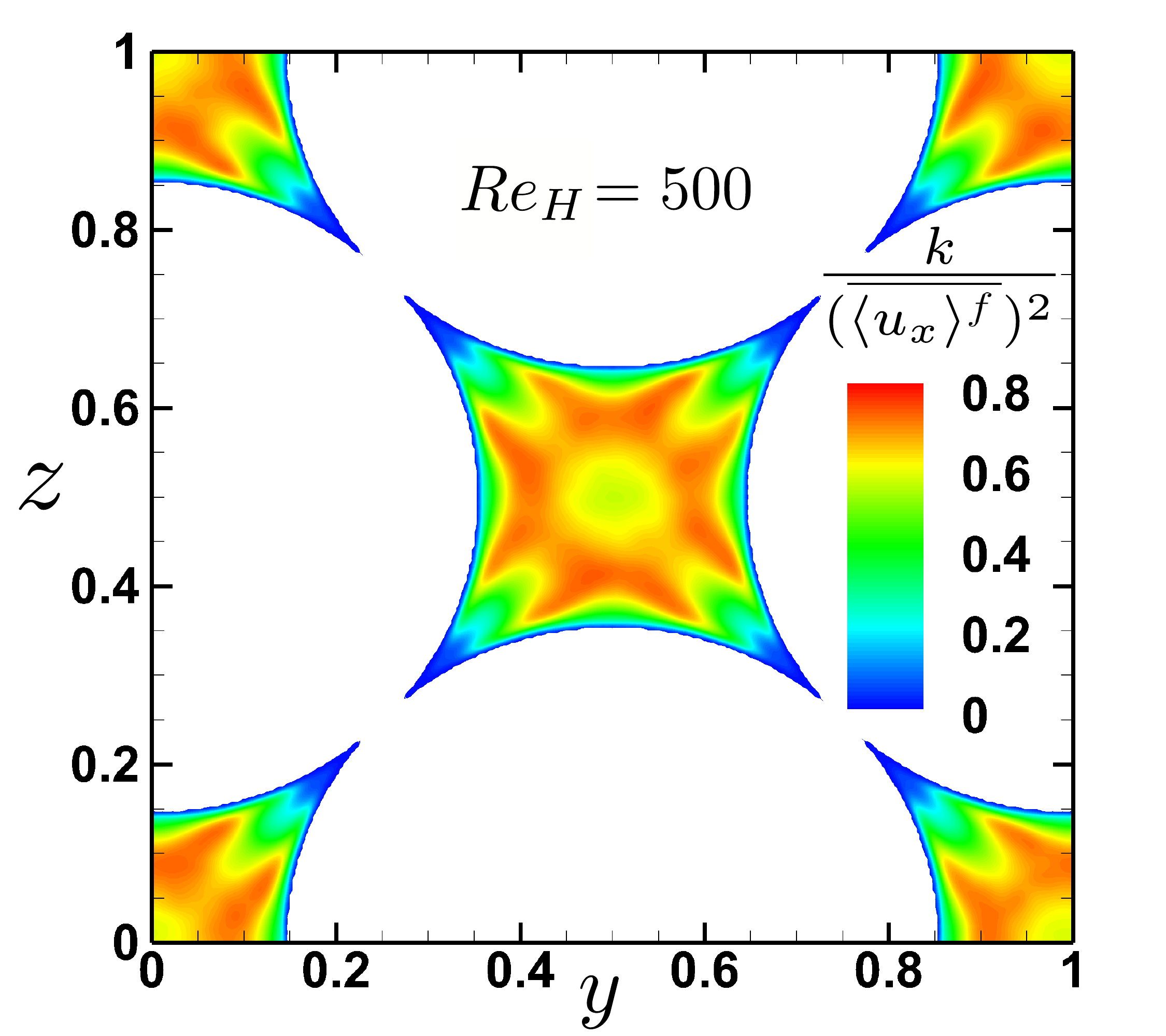}
        \caption{~}
        \label{fig_tke500_yz}
        \end{subfigure}
    \caption{Visualization of the mean velocity magnitude (top) and turbulent kinetic energy (bottom) at $Re_H=500$: (a--c)  $xy-$plane, and (b--d) $yz-$plane.}\label{fig_utke}
	\end{figure}

In order to get a good understanding of the simulated flow topology, distributions of the mean velocity magnitude $\overline U_m$ and the turbulent kinetic energy are presented first~(see figure~\ref{fig_utke}). The mean velocity $\overline U_m$ is calculated as $\sqrt{\overline {u_x}^2 + \overline {u_y}^2 +\overline {u_z}^2}$ and normalized by the mean interstitial velocity $\overline{\langle u_x \rangle^{f}}$. Two center slices are chosen as representative sections for visualizations. One is the center $xy-$plane, where the mean flow is going from left to right; the other the center $yz-$plane, where the mean flow in going into the slice. 

The mean flow enters the center pore from the left-hand side corners (regions labeled by (i$_1$) and (i$_2$) in figure~\ref{fig_utke}a) and accelerates to region (ii) due to the geometric constrictions; then the mean flow at the center starts to decelerate as it encounters the particle at region (iii), similar to an impinging jet. The flow then accelerates in the spanwise directions and leaves the pore through the right-hand side corners (regions (iv$_1$) and (iv$_2$)). The pattern of mean velocity distribution indicates how the mean flow in the pore is affected by the geometry. High Reynolds number flow features such as wake and large scale vortex shedding in an external flow behind a spherical particle, are not observed in this low porosity bed. A weak wake-like structure (small negative mean velocity just behind the left-hand side sphere near region (ii) in figure~\ref{fig_utke}a) is present, but is confined to a small region. The closely packed solid beads in the low porosity FCC configuration tend to break down large scale flow structures and prevent the generation of a significant wake region. Figure~\ref{fig_utke}b shows the distributions of the mean velocity magnitude on the center $yz-$plane. On this slice, the mean flow is moving perpendicularly into the page and there is a distinguished region with high velocity magnitude near each corner of the center pore.

The turbulent kinetic energy $k$, defined as ${\frac{1}{2} \left({\overline {{{{u'_x}}^2}}+\overline{{{u'_y}}^2}+\overline{{{u'_z}}^2}}\right) }$, is normalized by the square of the mean interstitial velocity $({\overline{\langle u_x\rangle^f}})^2$. Here the temporal fluctuation velocity $u_x'$ is defined as $ u_x -\overline{u_x}$. Figure~\ref{fig_tke500} and \ref{fig_tke500_yz} illustrate the distribution of normalized TKE on the center $xy-$ and $yz-$ planes, respectively. The overall magnitude of the TKE in the center pore region remains substantially high. It also suggests that the flow through the pore, although bounded by curved particle walls, is different from simple channel/duct flows even with complex boundaries~\citep{orlandi2018dns} 
wherein the TKE reaches a peak value near the boundaries and then decreases in the center region. The normalized TKE distributions on the center $yz-$plane presented in figure~\ref{fig_tke500_yz} shows some similarities with flow through a duct. Away from the wall, the TKE increases and reaches a peak value; and then decreases as the core region of the pore is approached. The homogeneous particle packing involving four particles aligned together seem to form a locally duct-like flow pattern in this section. However, it disappears in other sections away from the center $yz-$plane, owing to the three-dimensional nature of the spherical particles.
	
\begin{figure}
     \centering
      \begin{subfigure}[b]{0.3\textwidth}
            \includegraphics[height=\textwidth]{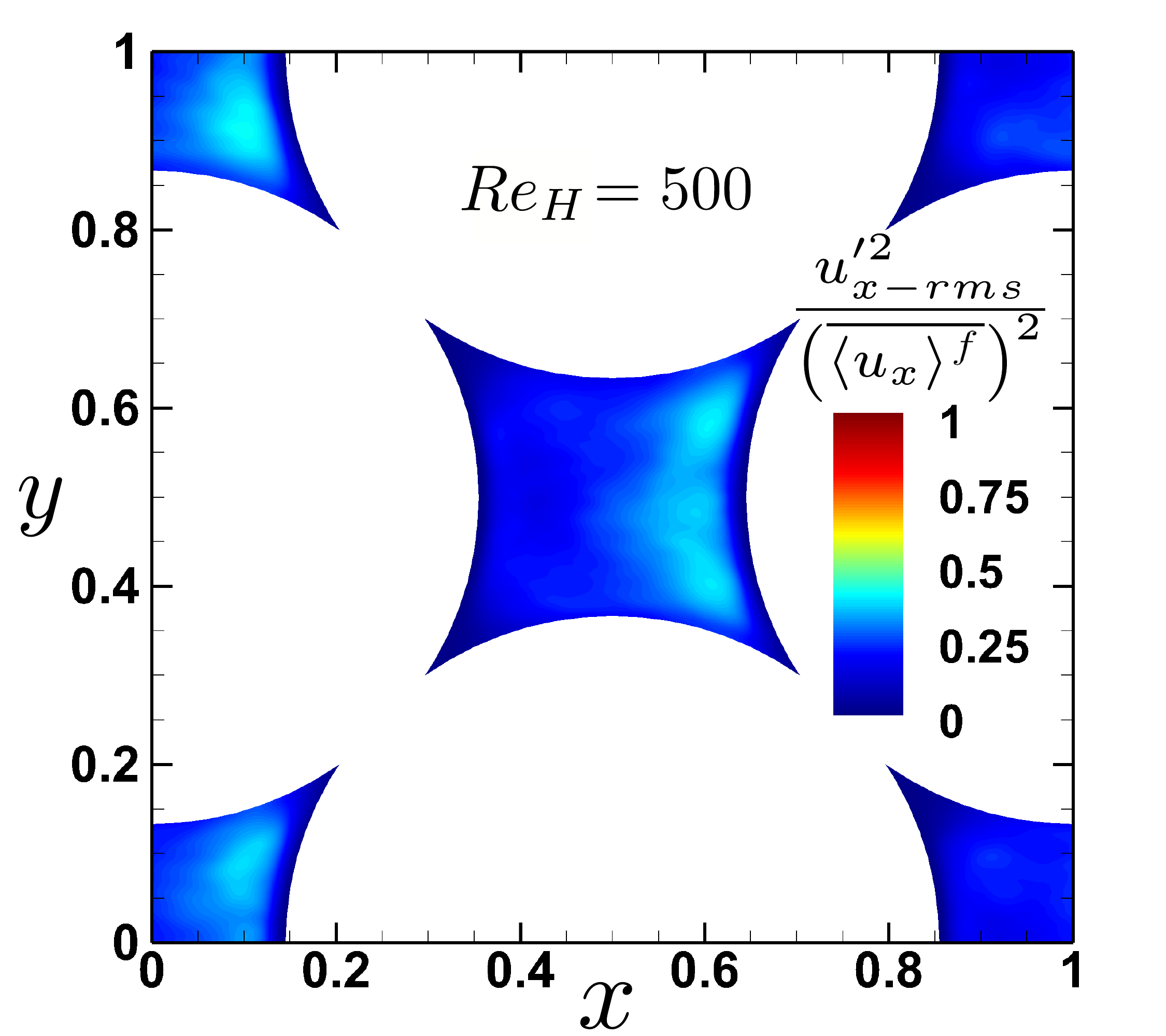}
            \caption{~}
                    \label{fig_500_ti_x}
      \end{subfigure}
       \begin{subfigure}[b]{0.3\textwidth}
             \includegraphics[height=\textwidth]{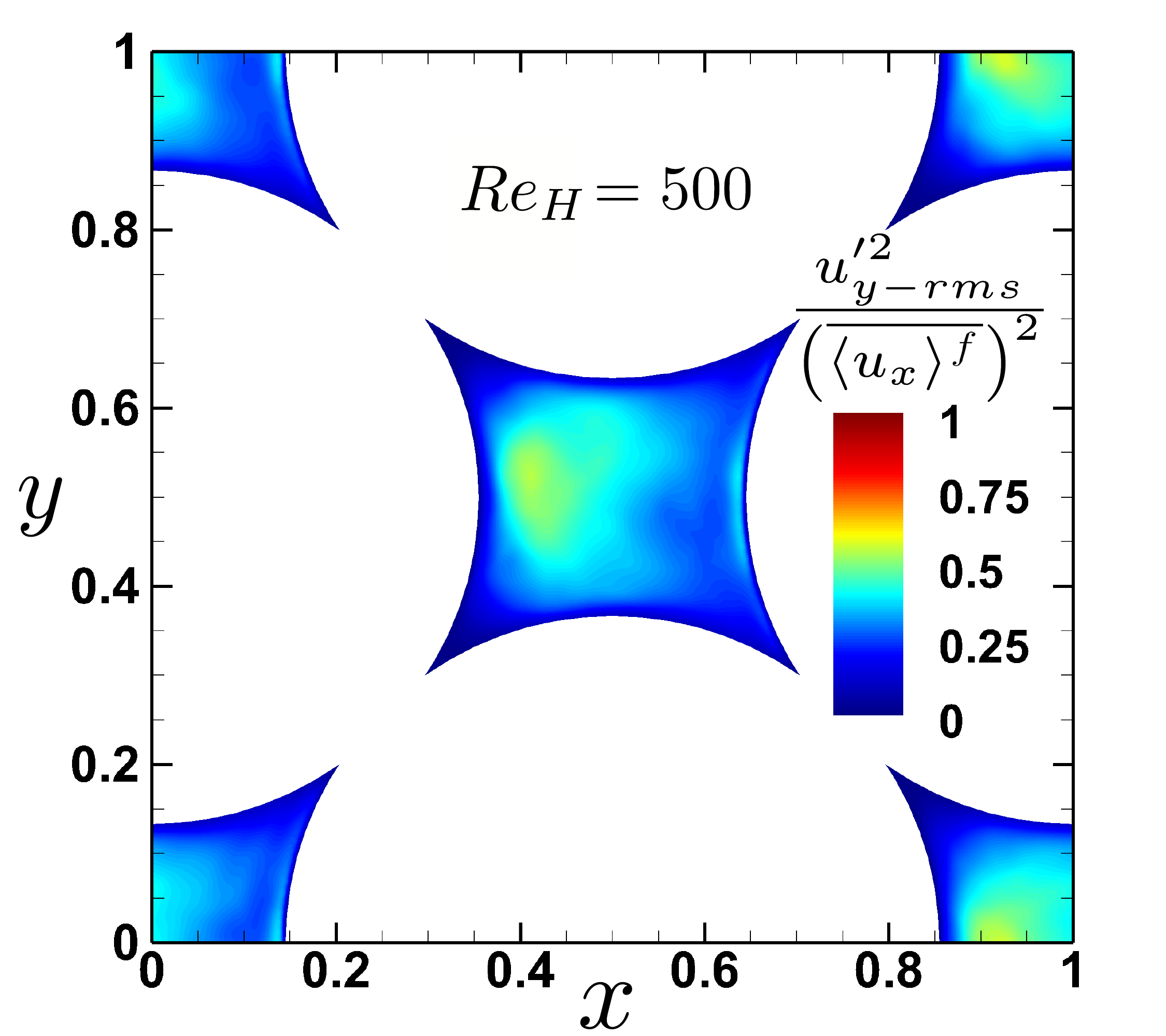}
            \caption{~}
             \label{fig_500_ti_y}
          \end{subfigure}
      \begin{subfigure}[b]{0.3\textwidth}
         \includegraphics[height=\textwidth]{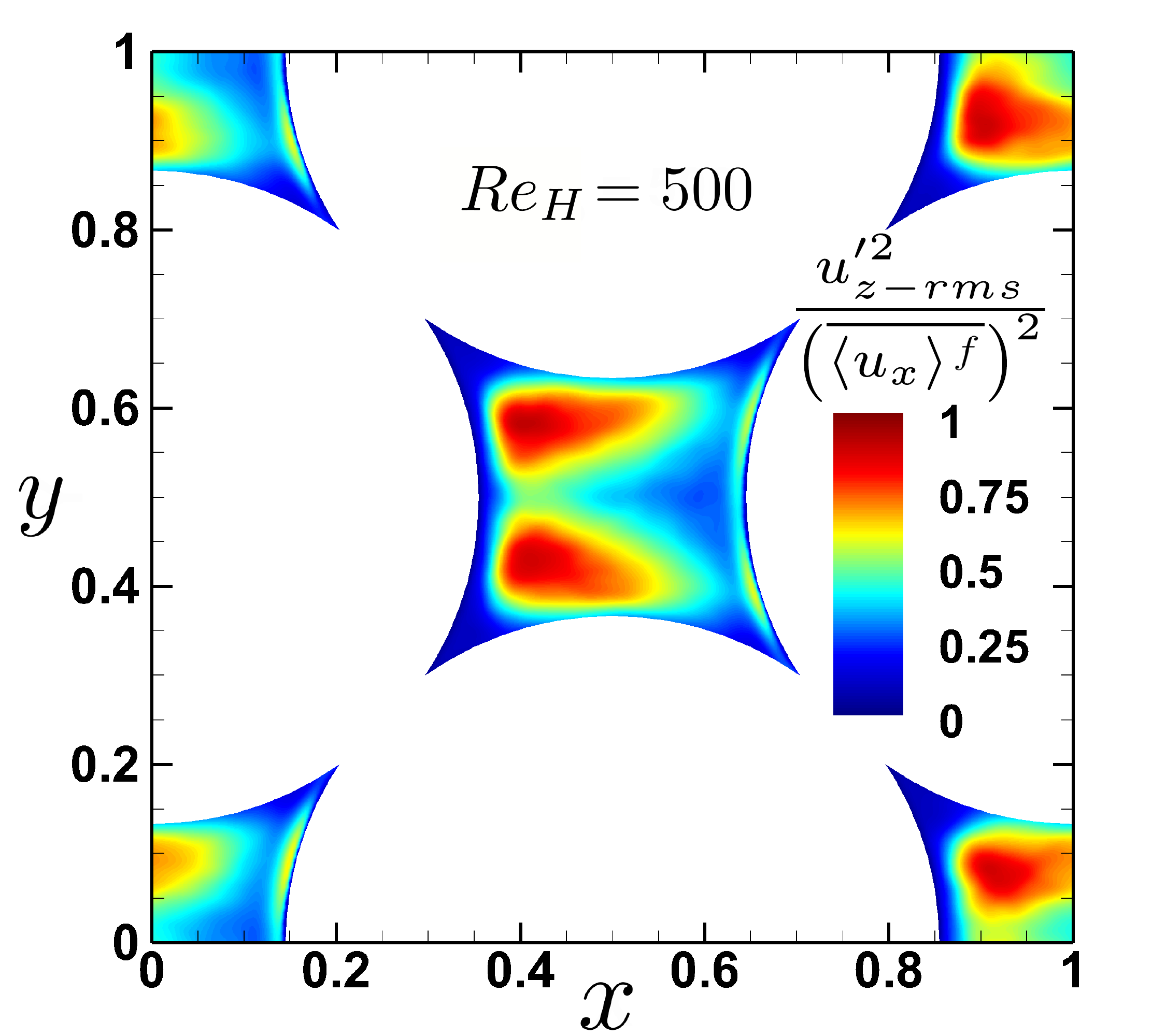}
            \caption{~}
             \label{fig_500_ti_z}
       \end{subfigure}
	\caption{Contours of turbulence intensity components normalized by the square of interstitial velocity on the center $xy-$plane: (a) $x$-component, (b) $y$-component, and (c) $z$-component. Mean flow is from left to right.}\label{fig_TI}
\end{figure}	

The turbulence intensity components in streamwise and spanwise directions are estimated to illustrate the anisotropy caused by the confined geometry, they are shown in the $xy-$plane in figure~\ref{fig_TI}. The $x$-component rms velocity, $ u'_{x-rms}$ is computed by
\begin{equation}
	u'_{x-rms} = \sqrt{\overline{u_x^2}-{\overline {u}_x}^2},
\end{equation}
likewise for $ u'_{y-rms}$ and $ u'_{z-rms}$.
The $x$-component turbulence intensity is only heavily pronounced in the region close to the right-hand-side particle (region (i) in figure~\ref{fig_500_ti_x}), which is mostly attributed to the impingement-like flow in this region. In this center $xy-$plane, the $x$-component of turbulence intensity, corresponding to the main flow direction, is weaker compared to the other two components. The maximum value of $x$-component turbulence intensity is about 80\% and 50\% of that for $y$- and $z$-components, respectively.  by~\citet{patil2013}. On the center $xy-$plane, these two components are distributed in various patterns. The $y$-component mostly concentrates near the left-hand-side particle in figure~\ref{fig_500_ti_y}. However for the $z$-component, the distribution is substantially different. There are two distinct regions with high magnitude of $z$-component turbulence intensity in figure~\ref{fig_500_ti_z} near the left-hand-side of the pore. More importantly, the shape of such regions is almost identical to the high TKE regions illustrated in figures~\ref{fig_tke500}, indicating that the $z$-component turbulence intensity has the most important contribution to total TKE at this particular plane. Although not shown here, it is noteworthy that, the distributions of $y$- and $z$-components of turbulence intensity on the center $xy-$plane are interchanged on the center $xz-$plane, due to the homogeneous and symmetric geometry of the FCC packing; and the $x$-component distribution remains the same.

To investigate the overall characteristics of both the streamwise and spanwise components of turbulence intensity and for purposes of comparison, intrinsic spatial average of these rms velocities (normalized by the square of intrinsic averaged velocity) was computed. Because of the periodicity and symmetry in the computational domain, the volume-averaged rms velocity in $z$-direction is similar to that in $y$-direction and hence not included here.
The turbulence intensity in both streamwise ($x$-) and spanwise ($y$-)directions are increasing with Reynolds number, as expected. However, the rms velocity in the spanwise direction ($y$- or $z$-) has a larger magnitude (0.288) than that in the $x$-direction (0.222). This behavior is quite different from the well-studied turbulent channel or duct flows even with complex boundary shapes~\citep{orlandi2018dns}, wherein the rms velocities in the wall normal directions are much smaller than the streamwise component. This again is caused by the three-dimensional effect of the complex configuration of the packed spheres, and the tortuous mean flow patterns within the pore. Qualitatively, a similar behavior that larger values of turbulence intensities in non-streamwise direction has also been observed experimentally in randomly-packed porous media~\citep{patil2013}.

Finally, the Eulerian and Lagrangian auto-correlations are used to compute the integral length and time scales, respectively.
To compute the Lagrangian auto-correlations, fluid tracer particles are tracked to obtain the Lagrangian trajectories. The Lagrangian auto-correlations are then computed according to Eq.~(\ref{eq:lag_auto}) (Ref.~\cite{monin1965}),
	\begin{align}
		\begin{aligned}
			\rho^L_{ij} (\tau) = \frac{{ {\langle {v}'_i({ X_0}, t) \:\:  {v}'_j({ X_0} ,t+\tau) \rangle} }}{{ {\Big[ \langle {{v_i}'^2}({ X_0}, t)\rangle\:\langle {{v_j}'^2}({ X_0}, t+\tau)\rangle \Big]^{1/2}} }}
			\label{eq:lag_auto}
		\end{aligned}
	\end{align}
where $\rho^L_{ij}$ is the Lagrangian auto-correlation, $v'_i$ the $i$-th component of the particle fluctuation velocity and $\langle \cdot  \rangle$ represents ensemble averaging. The Lagrangian integral time scale, $T_{11}^{L}$ is simply given by the integral of the auto-correlation function.

The Eulerian integral length scale ($L_{11}^E$) normalized by the bead diameter ($D_B$) is found to be $0.0884$, whereas the Lagrangian integral time scale normalized by the flow time scale ($T_{11}^L$) based on the interstitial velocity ($U_{int}$) and the bead diameter ($D_B$) is about 0.356 for the present flow conditions. The integral length scale is only about 10\% of the sphere diameter, indicating that the coherent structures are confined within the pore. Such observation supports the pore scale prevalence hypothesis (PSPH) and the results reported in \citet{jin2015}. This implies that, the turbulence in the pore-scale is strongly affected by the porosity, and restrained by the pore size. The integral time scale is also smaller than the flow time scale, suggesting that the Lagrangian coherent structures are restricted by the pore size. As a result, the single periodic unit cell considered in the present work is sufficient.
The overall dissipation rate $\langle\epsilon\rangle$ was estimated from the TKE budget first. Then the Kolmogorov time scale is computed as $\tau_\eta =\sqrt{{\nu}/{\langle\epsilon\rangle}}$.
For the case $Re_H=500$, $Re_\lambda$ can be estimated to be around 32, which is computed as the drape of the fitted parabola to the Eulerian auto-correlation.
Assuming isotropic turbulence, which is not the case for the flow in porous media, and using the same definition as in~\cite{matsuda2020scale}, an even smaller value of $Re_\lambda \sim 21$ is obtained for the Taylor microscale Reynolds number.

\subsection{Inertial particle dynamics and clustering} 
\label{sec:inertial_part}
\begin{figure}
     \centering
      \begin{subfigure}[b]{0.48\textwidth}
            \includegraphics[height=\textwidth]{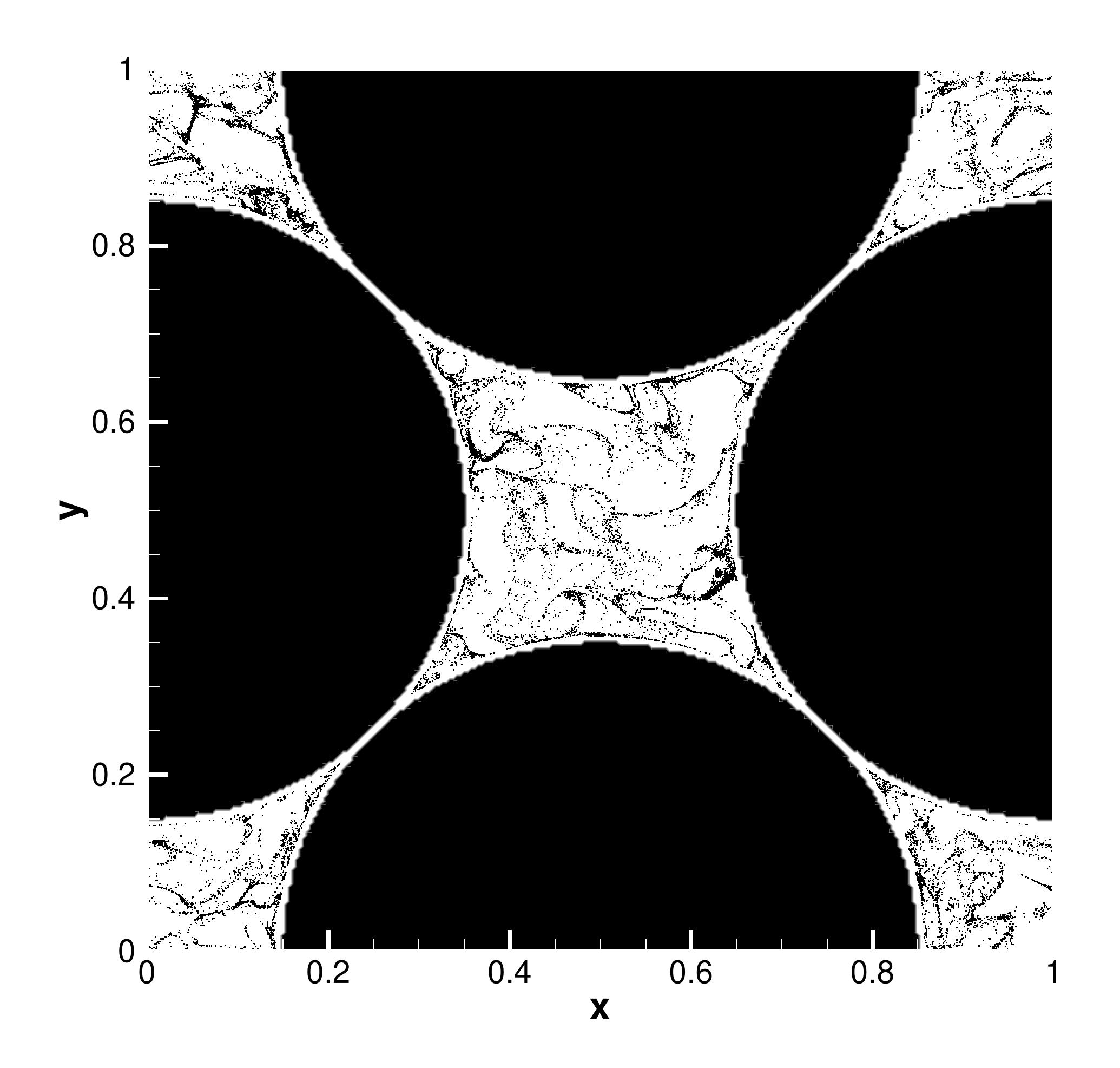}
            \caption{~}
                    \label{St1lam}
      \end{subfigure}
       \begin{subfigure}[b]{0.48\textwidth}
             \includegraphics[height=\textwidth]{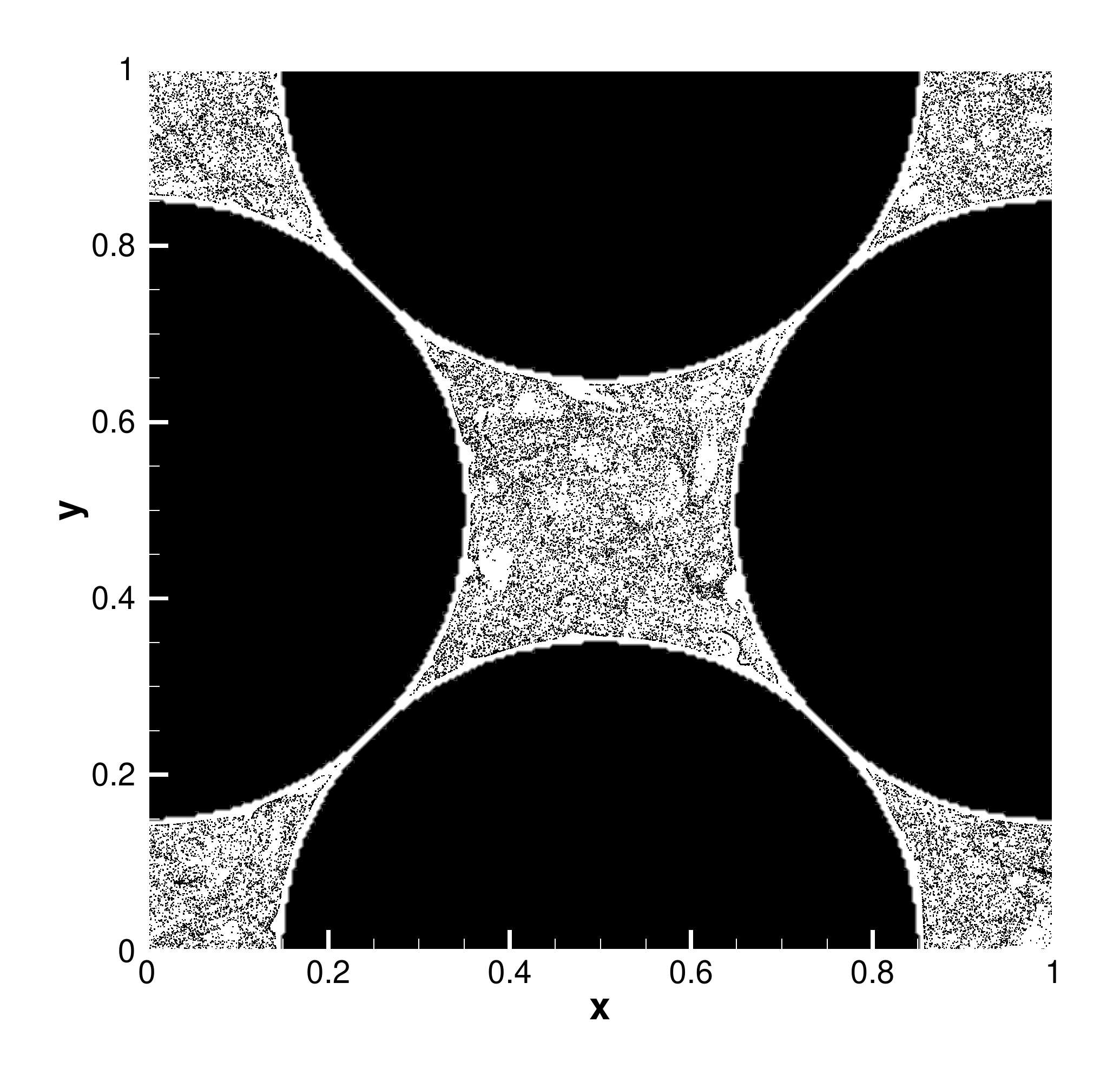}
            \caption{~}
             \label{St0.1lam}
          \end{subfigure}
      \begin{subfigure}[b]{0.48\textwidth}
         \includegraphics[height=\textwidth]{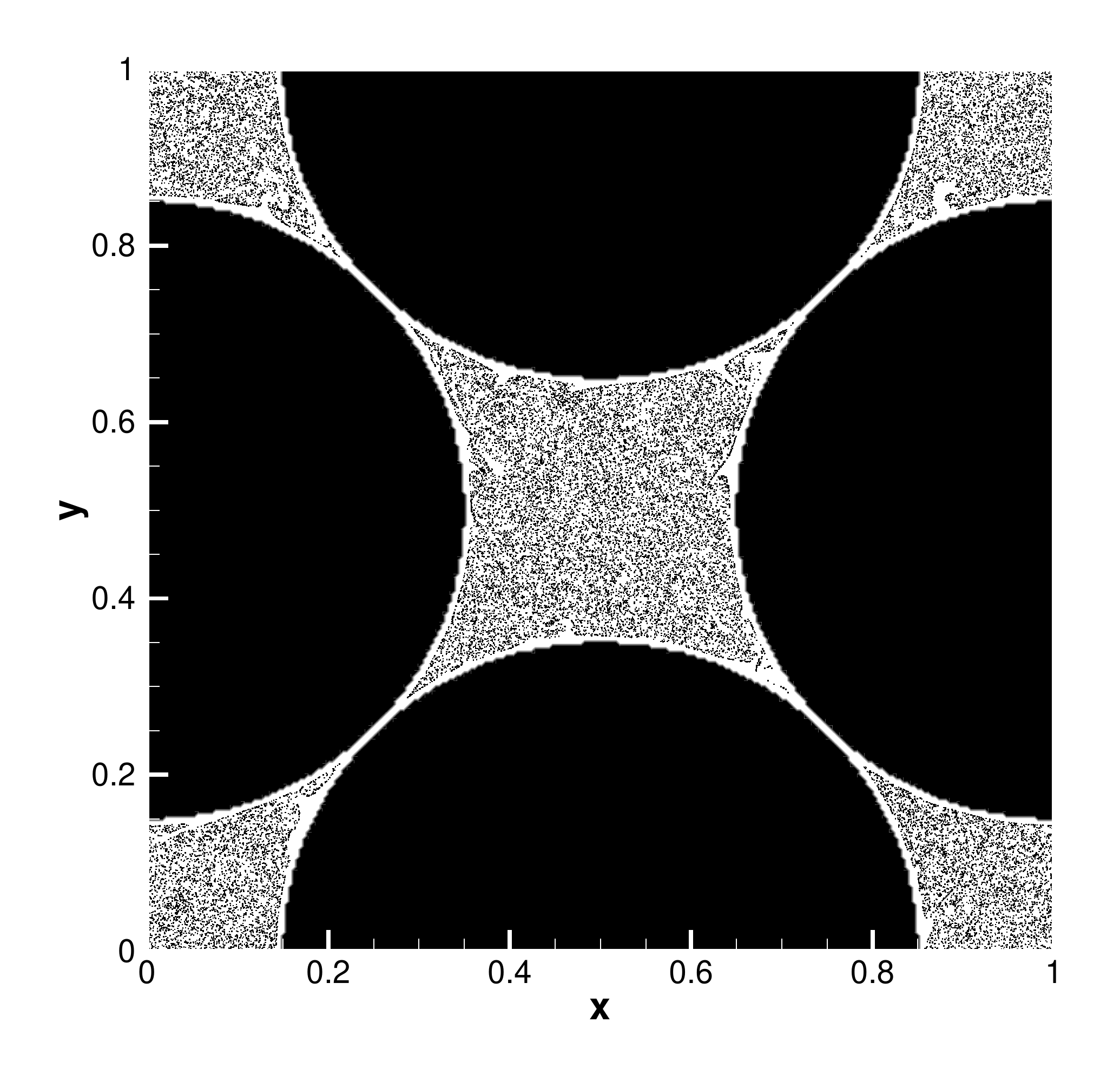}
            \caption{~}
       \end{subfigure}
             \begin{subfigure}[b]{0.48\textwidth}
         \includegraphics[height=\textwidth]{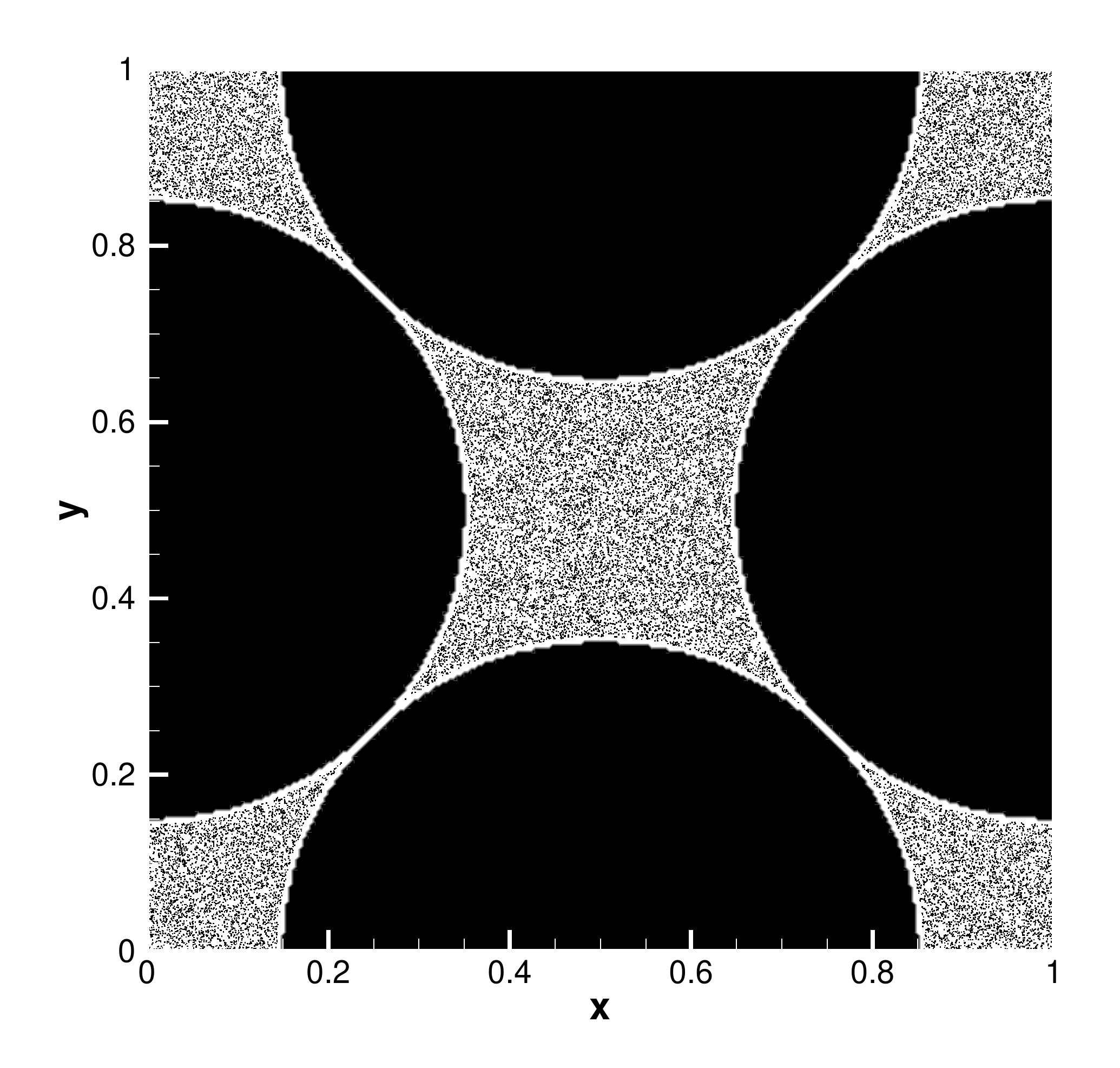}%
\caption{~}
     \end{subfigure}
	\caption{Instantaneous distribution of particles in the $xy-$plane: (a) $St=1$, (b) $St=0.1$, (c) $St=0.01$, and (d) uniform random distribution. }\label{fig_lam}
\end{figure}

Figure~\ref{fig_lam} shows the instantaneous distribution of particles for three different Stokes numbers in the $xy-$plane after a stationary state is reached. Uniform random distribution of particles in the porous geometry, representative of fluid particles in the limit of $St\rightarrow0$, is also shown for comparison. Significant particle clustering is observed for $St=1$ as expected, whereas for $St=0.01$ particles are more uniformly distributed with only few pockets of voids and clusters near the bead boundaries. The clustering of inertial particles and effect of the bead walls is evaluated by conducting a multiscale analysis of the particle number density.

\subsubsection{Voronoi tessellation for particle clustering}
\label{sec:voronoi}
Voronoi tessellation, see e.g.~\citet{aurenhammer1991voronoi}, is a technique to construct a decomposition of the fluid domain, into a finite number of Voronoi cells. If there are finite number of points (particles) dispersed in space, a Voronoi cell, is defined as a region of all points that are closer to a particle than any other particles. The volume of the Voronoi cell is referred to as the Voronoi volume, $V_p$. The magnitude of this volume can be used to quantify particle clustering and void regions in a three-dimensional space, smaller volume indicating pronounced clustering. Three-dimensional Voronoi tessellation is applied to the particle data obtained from the present DNS data using the Quickhull algorithm provided by the Qhull library in python~\citep{barber1996quickhull}, which has a computational complexity of ${\mathcal O}(N_p {\rm log}N_p)$.

\begin{figure}
\centering
\includegraphics[width=0.95\linewidth]{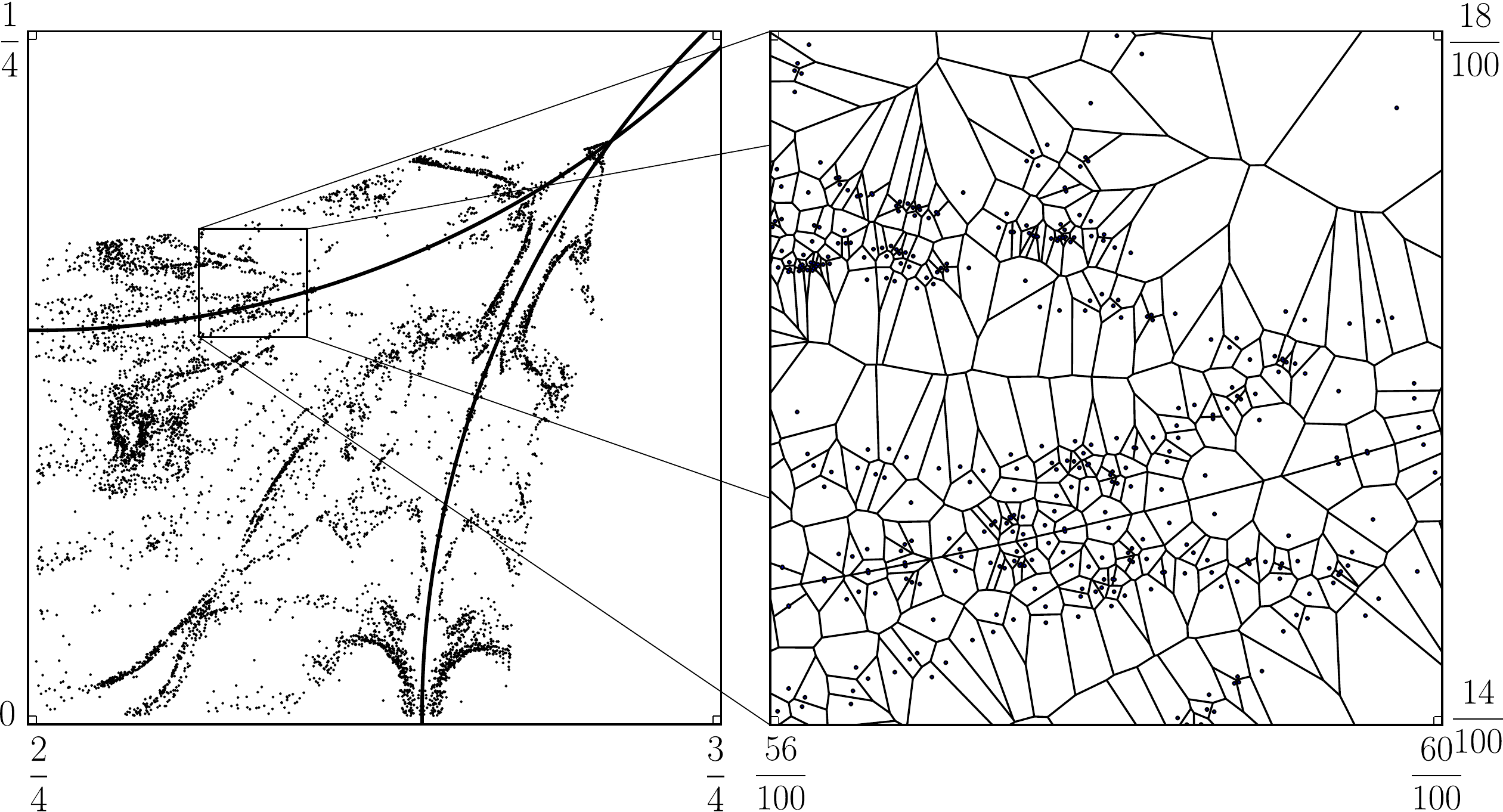}
\caption{ Particle distribution for $St = 1$ in a slice of thickness $1/100$ with spheres and mirror particles (left). A magnified view with Voronoi tessellation (right). 
}
\label{fig:Voronoi_tessellation_mirror}
\end{figure}

\begin{figure}
     \centering
            \includegraphics[height=0.6\textwidth]{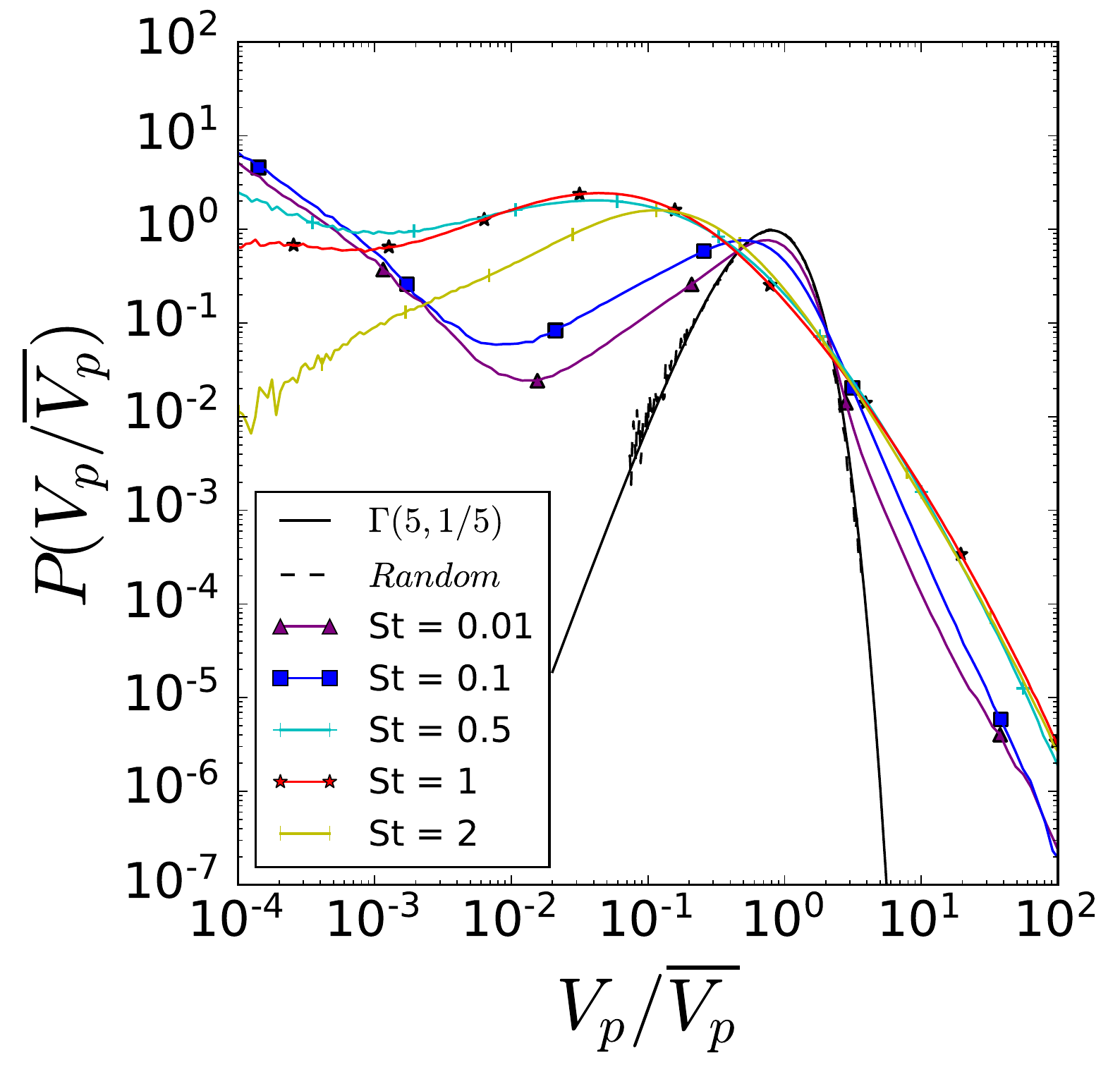}
	\caption{PDF of volumes of Voronoi cells in log-log representation normalized by the mean volume for different Stokes numbers as well as for particles distributed randomly following a Poisson distribution. The PDF of randomly distributed particles is also compared with the gamma distribution and shows perfect agreement.}
    \label{fig:Volume_Voronoi}
\end{figure}
To construct the Voronoi tessellation of particles in the presence of the embedded bead boundaries, a special treatment is needed for particles near the bead walls. Figure~\ref{fig:Voronoi_tessellation_mirror} shows how the geometry of the spherical beads is taken into account. All particles located at a given small distance from each sphere boundary is assigned a mirror particle. Introducing these ghost particles then accounts for the bead boundary, and the Voronoi tessellation can then be constructed making use of the mirror particles. To verify that this approach works, a random distribution of particles in the fluid domain is first considered. For randomly distributed particles, the PDF of the Voronoi volume becomes a gamma distribution~\citep{ferenc2007size}. For the 3D case, the PDF of Voronoi volume is given by $\Gamma(5,1/5)$, where $\Gamma(k,\theta)$ corresponds to the gamma distribution,
\begin{equation}
    f_{V_p}(x) = \Gamma(k)^{-1}\theta^{-k}x^{k-1}{\rm exp}(-x/\theta),
\end{equation}
where $k$ and $\theta$ are shape parameters of the PDF, respectively.
Figure~\ref{fig:Volume_Voronoi} shows the PDF of the Voronoi volume $V_p$, normalized by the mean volume $\overline{V_p}$, for different Stokes numbers as well as for randomly distributed particles. The PDF for random distribution follows closely with
the gamma distribution, as expected. It should be noted that in the present fictitious domain method, the bead boundaries are smeared over a grid control volume owing to the interpolation function between the bead material points and the computational grid typical of penalization based techniques. Accordingly, to perfectly match the random particle PDF to the gamma distribution, the sphere radius had to be increased by about 1.1\% ($0.35749$ instead of $0.353553$) which is comparable to the grid resolution used. This small modification in sphere geometry is applied to all particle distributions obtained from different Stokes numbers. 

Figure~\ref{fig:Volume_Voronoi} 
shows that the PDFs of Voronoi volumes for different Stokes numbers intersect with the gamma distribution, i.e. the one for the random particles. The observed behavior is similar to that found for homogeneous isotropic turbulence (HIT), see~\citep{oujia2020jfm}, but with some key differences. The number of large Voronoi cells increases with increasing Stokes numbers and then stabilizes. This behavior is resembling the HIT case. It is also seen that as the Stokes number increases and gets closer to $1$, the number of small normalized volumes ($10^{-2}$--$5 \cdot 10^{-1}$) 
also increases similar to the HIT case. However, large number of very small volumes ($10^{-4}$--$10^{-2}$) are observed for nearly all Stokes numbers, even for very small Stokes numbers. Such a behavior was not observed in HIT, wherein the number of very small Voronoi volumes would decrease monotonically. This is attributed to the interaction of particles with the bead surfaces in the present case. Particles with finite, non-zero Stokes numbers interact with the bead surface and undergo specular reflection. In addition, as the particles approach the bead surfaces, their velocities are slowed down significantly as the fluid velocity itself is smaller owing to the no-slip condition. Thus, existence of large number of very small volumes is mainly attributed to the collision of the particles with the bead surfaces. Note that, fluid particles (or tracers) would not collide with the bead surfaces, except at the stagnation points and at large times. Thus, the distribution of fluid tracer particles will follow the one of random particles, i.e. the gamma distribution.

\subsubsection{Voronoi-based divergence of particle velocity}
\label{subsubsec:divergence}
To understand the clustering dynamics of inertial particles, 
the particle number density $n$, as a continuous function, is commonly used~\citep{oujia2020jfm}. It satisfies the conservation equation,
\begin{equation}
    \frac{D}{Dt}(n) = \frac{\partial n}{\partial t} + {\mathbf u}_p\cdot \nabla n = - n \, \nabla\cdot {\mathbf u}_p \, .
    \label{eq:cons_particlenumber}
\end{equation}
The divergence of the particle velocity ${\mathbf u}_p$ appears as a source term in the number density equation. However, finding the divergence of the particle velocity is not straightforward as the discrete particle distribution is not continuous, and the particle velocity is only known at the discrete particle locations, 
${\mathbf u}_{p,j} = {\mathbf u}({\mathbf x}_{p,j})$, 
but not everywhere else. Moreover it can be multi-valued.~\citet{oujia2020jfm} proposed a method to compute the divergence of the particle velocity ${\mathcal D} = \nabla \cdot {\bf u}_p$ in a discrete manner using a Lagrangian approach. From the conservation equation for the particle number density (eq.~\ref{eq:cons_particlenumber}), one obtains, ${\mathcal D} = -\frac{1}{n}\frac{Dn}{Dt}$.

To calculate the Lagrangian derivative of $n$,~\citet{oujia2020jfm} defined the local number density $n_p$ as the number density averaged over a Voronoi cell, which is given by the inverse of the Voronoi volume $V_p$; i.e. $n_p =1/V_p$. Then it was shown that the divergence is obtained to the first-order approximation as,
\begin{equation}
    {\cal D}_p = \frac{2}{\Delta t} \, \frac{V^{k+1}_p-V^{k}_p}{V^{k+1}_p+V^{k}_p} + O(\Delta t) \; ,
\end{equation}
where $V^{k}_p$ denotes the Voronoi volume at time instant $t^k$.
This shows that the divergence of the particle velocity can be estimated from subsequent Voronoi volumes, provided the time step is sufficiently small and the number of particles is sufficiently large. To obtain the subsequent Voronoi volumes, the particle positions were linearly advanced by ${\mathbf u}_p$; i.e, ${\mathbf x}_p^{k+1}= {\mathbf x}_p^{k}+{\mathbf u}_p\Delta t$. The time-step was set to be same as the flow solver time step, which is sufficiently small for the present DNS study.
The influence of the step size has been checked and found to give same result as in \citet{oujia2020jfm}.
For the different time steps, the same probability distribution of the divergence was obtained, except that the extreme values of the divergence are changing with $\Delta t$.
\begin{figure}
\centering
             \includegraphics[height=0.6\textwidth]{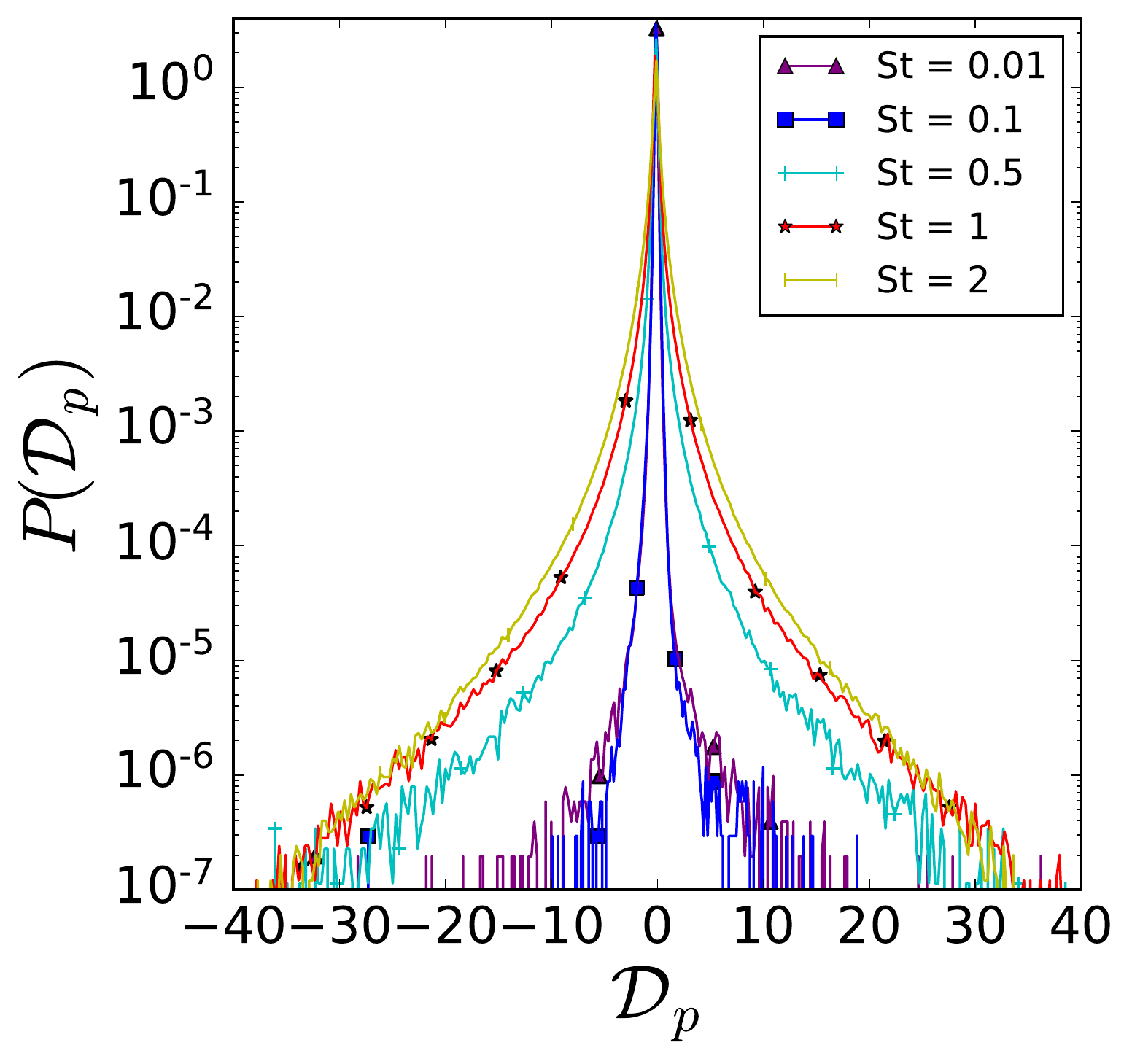}
	\caption{PDF of divergence of particle velocity for different Stokes numbers.}\label{fig:Divergence_Voronoi}
\end{figure}

Determining the divergence with the above Lagrangian approach requires two time instants of the Voronoi volumes. Above it was shown that the bead geometry, i.e. the presence of curved walls, necessitates some special treatment for computing the Voronoi tesselation. Mirror particles are introduced to account for the geometry. For computing the discrete divergence the particles are then linearly advanced in time to obtain the volume at a subsequent time instant. This procedure implies that some particles could enter the beads and the flow geometry is not respected anymore. Consequently the volume of adjacent cells are impacted and the computed divergence value is erroneous. To remove this artifact, particles having a distance less than $0.35749$ from each of the beads centers 
were not taken into account.
This value has been determined by considering the statistics of the divergence as a function of the wall distance. While for the mean value, which is close to zero, no significant influence was found, for the higher order statistics (variance, skewness and flatness), a significant change with the distance, even by orders of magnitude, was observed. For wall distances larger than $0.35749$, the values were found to be stable and remained almost constant.

%


The PDFs of the Voronoi-based divergence for different Stokes numbers are shown in figure~\ref{fig:Divergence_Voronoi}. They are centered around 0 and confined between $-40$ and $40$.
It can be observe that extrema  correspond to $\pm 2/\Delta t$ 
which is an upper/lower bound, even a rigorous bound when neglecting $O(\Delta t)$ terms, similar to what is found for HIT~\citep{oujia2020jfm}. 
The divergence should become closer to zero as the Stokes number decreases to zero because the fluid particles in an incompressible flow are divrgence free. However, in figure~\ref{fig:Divergence_Voronoi}, the divergence for $St=0.01$ is comparable to that of $St=0.1$.
It can be deduced that 
the divergence for these Stoks numbers is mainly caused by a geometrical effect due to Voronoi tessellation, which is also discussed in  \citet{oujia2020jfm}. 
However for larger Stokes numbers, physical effects do predominate.
Table \ref{table:stat_vor_div} assembles the variance and flatness of the divergence ${\cal D}_p$ as a function of the Stokes number. 
The flatness decreases as the Stokes number increases (with the exception of $St=0.1$), 
this implies that the tails of the PDFs decay faster as the Stokes number increases.
In contrast the variance has the opposite behavior and thus the PDFs are becoming wider and wider for increasing Stokes number, confirmed in figure~\ref{fig:Divergence_Voronoi}.
Thereafter the Voronoi analysis will be discussed only for Stokes numbers larger or equal to $0.5$, to avoid the influence of the geometrical effect. 
%

\begin{table}
\begin{center} 
\begin{tabular}{ lcccccccc } 

 St & 0.01 & 0.1  & 0.5 & 1 & 2\\ 
 Variance & 0.0083 & 0.0093 & 0.0952 & 0.2796 & 0.5423 \\ 
 Flatness & 5514.9 & 734.50 & 832.00 & 291.92 & 105.70 \\ 
\end{tabular}
\caption{Variance and flatness of divergence ${\cal D}_p$ as a function of the Stokes number. 
\label{table:stat_vor_div}}
\end{center}
\end{table}

\begin{figure}
\centering
\includegraphics[width=0.6\linewidth]{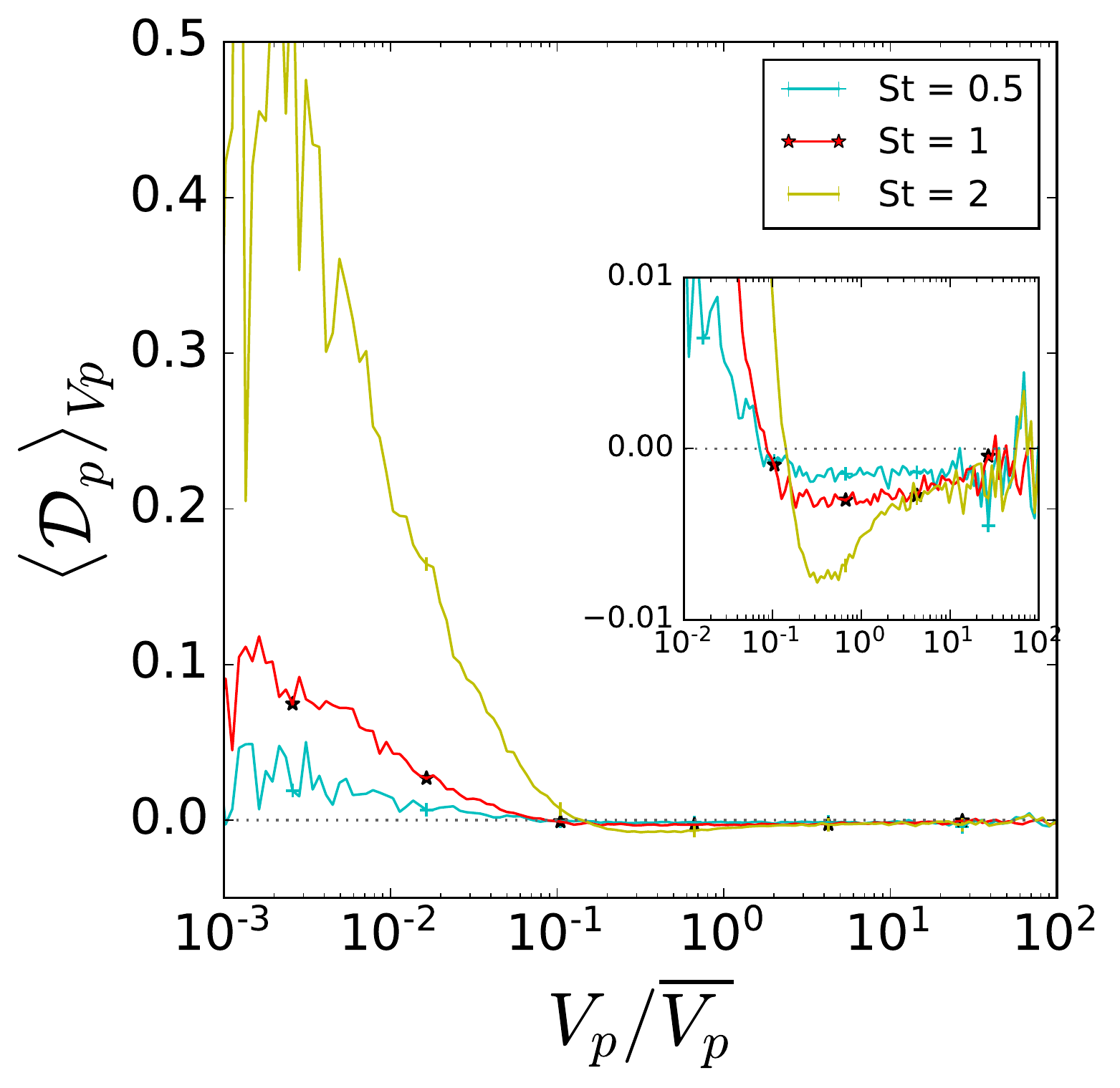}
\caption{Mean divergence $\langle {\cal D}_p \rangle_{Vp}$ as a function of the Voronoi volume for different Stokes numbers. }
\label{fig:Mean_div_vol}
\end{figure}

The mean of the divergence as a function of the volume is defined as 
\begin{equation}
    \langle {\cal D}_p \rangle_{Vp} = \frac{1}{P(V_p/\overline{V_p})}\int_{-\infty}^{+\infty} {\cal D}_p \, P({\cal D}_p,V_p/\overline{V_p}) \, d{\cal D}_p
    \label{eq:average_divergence},
\end{equation}
The values are shown in figure~\ref{fig:Mean_div_vol} for 3 Stokes numbers, $St=0.5, 1$ and $2$. The insert is a zoom to focus on the negative values. 
Recall that negative values correspond to convergence of the particles and positive values to divergence. 
It can be seen that the mean of the divergence is positive for small volumes, negative for large volumes and similarly to~\citet{oujia2020jfm} the amplitude increases with the Stokes number. This implies presence of cluster formation for large volumes, cluster destruction for small volumes and that these two behaviors are amplified when the Stokes number increases. 
The zero-crossing point is $V_p/\overline{V_p} \approx 0.06-0.2$, a value close to what is observed in \citet{oujia2020jfm}. 
%

In summary, the above findings for the Voronoi analysis confirm the results found in \citet{oujia2020jfm} for homogeneous isotropic turbulence, in particular the $St$ dependence of cluster formation and destruction. Differences are found for small Voronoi volumes below 0.01, they behave differently due to the influence of the bead geometry. The zero crossing point of the mean value of the divergence is likewise shifted towards smaller values.



\subsection{Eulerian field: Particle number density}
\label{sec:numden}
The number density of the discrete particle positions is obtained using a histogram method by binning particles onto an equidistant cubic grid of $N_g^3$ grid points~\citep{matsuda2020scale}. Irrespective of the grid resolution used in the direct numerical simulation, $N_g=2^8$ was used to calculate the number density as,
\begin{equation}
    \label{eq:numden}
    n({\mathbf x},t) = \sum_{i_1,i_2,i_3=0}^{N_g-1}\left\{\int_{\Omega}K_h({\mathbf x}_{i_1,i_2,i_3}-{\mathbf x}^{\prime})
    {1 \over n_0}\sum_{m=1}^{N_p}\delta({\mathbf x}^{\prime} - {\mathbf x}_{p,m}(t))
    d{\mathbf x}^{\prime}\right\}h^3 K_h({\mathbf x}-{\mathbf x}_{i_1,i_2,i_3}),
\end{equation}
%
where ${\mathbf x}_{i_1,i_2,i_3}=h(i_1+1/2,i_2+1/2,i_3+1/2)$ is the box position, and $K_h({\mathbf x}) = 1/h^3$ for $-h/2\leq x_i \leq h/2$ ($i=1,2,3)$, while $K_h({\mathbf x})=0$ otherwise, is a piecewise constant function, $h=L/N_g$, and $n_0 = N_p/L^3$ is the mean dimensional number density, where $N_p$ is the total number of particles and $L$ is the side length of the cubic domain. With non-dimensionalization by the mean number density, the above equation satisfies $\left<n\right>=1$. 

It should be noted that the entire computational domain volume is used to compute the number density, even though some part of the volume is a solid bead region to simplify the number density computation. The porosity of the unit cell can be used to relate this to the number density calculated based on the fluid volume only. The consequence of presence of solid beads within the computational domain is that, a uniform distribution of inertial particles in the fluid domain results in non-uniform number density variations across the bead surface. These gradients in number density, even for a uniform inertial particle distribution, can result in non-zero wavelet decomposition. To analyze the true clustering of inertial particles, this effect of pseudo variations in number density due to the bead geometry, need to be removed before performing the wavelet decomposition. Therefore  the number density gap from $n({\mathbf x})$ is subtracted as
\begin{equation}
    \label{eq:densitywavelet}
    n^\prime({\mathbf x}) = n({\mathbf x}) - \phi^{-1} \langle n \rangle~\left( 1- \chi({\mathbf x}) \right)
\end{equation}

with $\chi(\mathbf{x})$ the mask function :
\begin{equation}
\label{eq:chi}
\chi(\mathbf{x}) = \bigg\{
\begin{array}{l l}
  1 & \quad \text{if $\mathbf{x} \in \Omega_s$ }\\
  0 & \quad \text{if $\mathbf{x} \in \Omega_f$ }\\ 
 \end{array}
\end{equation}
where $\Omega_s$ is the solid domain, $\Omega_f$ the fluid domain and $\Omega=\Omega_f \cup \Omega_s$ the computational domain. For sake of clarity, $n^\prime({\mathbf x})$ is denoted by $n({\mathbf x})$.

%


\subsubsection{Scale dependent wavelet analysis of number density}
\label{sec:wavelet}
Inertial particle clustering and related multiscale statistics are quantified using the orthogonal wavelet decomposition~\citep{Mallat1999,daubechies1993ten} of the particle number density. Consider the particle number density, $n({\mathbf x},t)$, at a given instant $t$, within the computational domain of a triply periodic, $L^3$ cubic box. This scalar field is decomposed into a 3D orthogonal wavelet series to unfold into scale, positions and seven directions ($\mu = 1, 2, ...., 7$). The 3D mother wavelet, $\psi_{\mu}({\mathbf x})$, is hereby based on a tensor product construction and a
family of wavelets, $\psi_{\mu,{\bm \lambda}}({\mathbf x})$  can be generated by dilation and translation. 
an orthogonal basis of $L^2({\mathbb{R}}^3)$. 
The multi-index ${\bm \lambda}= (j,i_1, i_2, i_3)$ denotes the scale $2^{-j}$ and position $L\times 2^{-j}{\mathbf i} = L\times 2^{-j}(i_1,i_2,i_3)$ of the wavelets for each direction, where $i_{\ell} = 0,...,2^{j-1}~ (\ell=1,2,3)$. The wavelets are well-localized in space around position, $L\times 2^{-j}(i_1,i_2,i_3)$, and scale, $2^{-j}$, oscillating, and smooth. 
A periodization technique~\citep{Mallat1999} is applied to the wavelets.
The spatial average of $\psi_{\mu,{\bm \lambda}}({\mathbf x})$, defined by, $\left< \psi_{\mu,{\bm \lambda}} \right> = L^{-3}\int_{{\mathbb T}^3} \psi_{\mu,{\bm \lambda}}({\mathbf x}) d{\mathbf x}$ vanishes for each index, which is a necessary condition for
being a wavelet.

Similar to \citet{matsuda2020scale} the number density field $n({\mathbf x},t)$ sampled on $N_g^3 = 2^{3J}$ equidistant grid points, can be developed into an orthogonal wavelet series: 
\begin{equation}
n({\mathbf x}) = {\overline n}_{000}({\mathbf x}) + \sum_{{ j}=0}^{J-1} n_{j} ({\mathbf x}),
\end{equation}
where $n_{j}({\mathbf x})$ is the contribution of $n({\mathbf x})$ at scale $2^{-j}$ defined as,

\begin{figure}
     \centering
      \begin{subfigure}[!htpb!]{0.47\textwidth}
            \includegraphics[height=\textwidth]{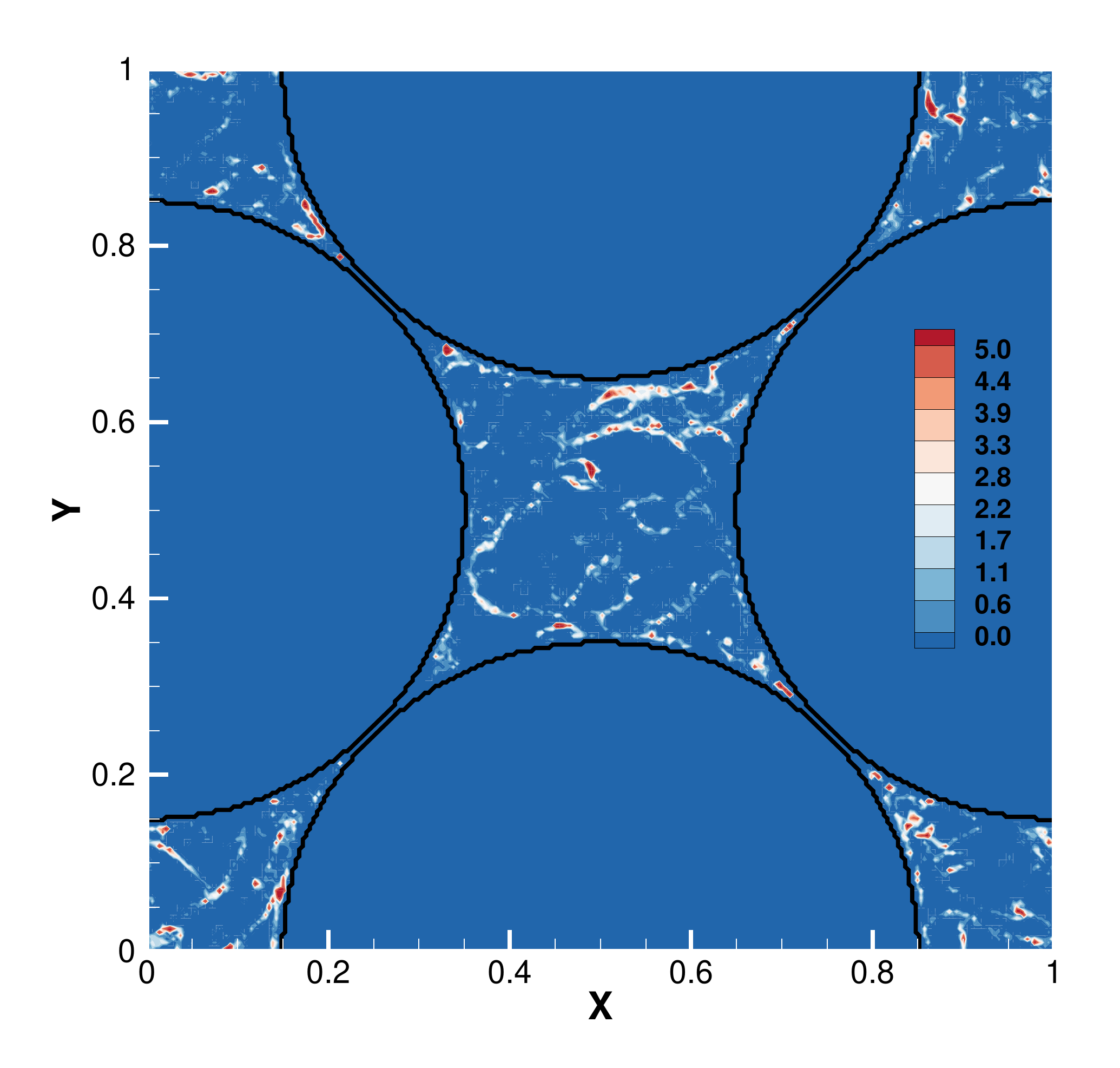}
            \caption{$n^\prime$, $St=1$}
                    \label{St1full}
      \end{subfigure}
       \begin{subfigure}[!htpb!]{0.47\textwidth}
             \includegraphics[height=\textwidth]{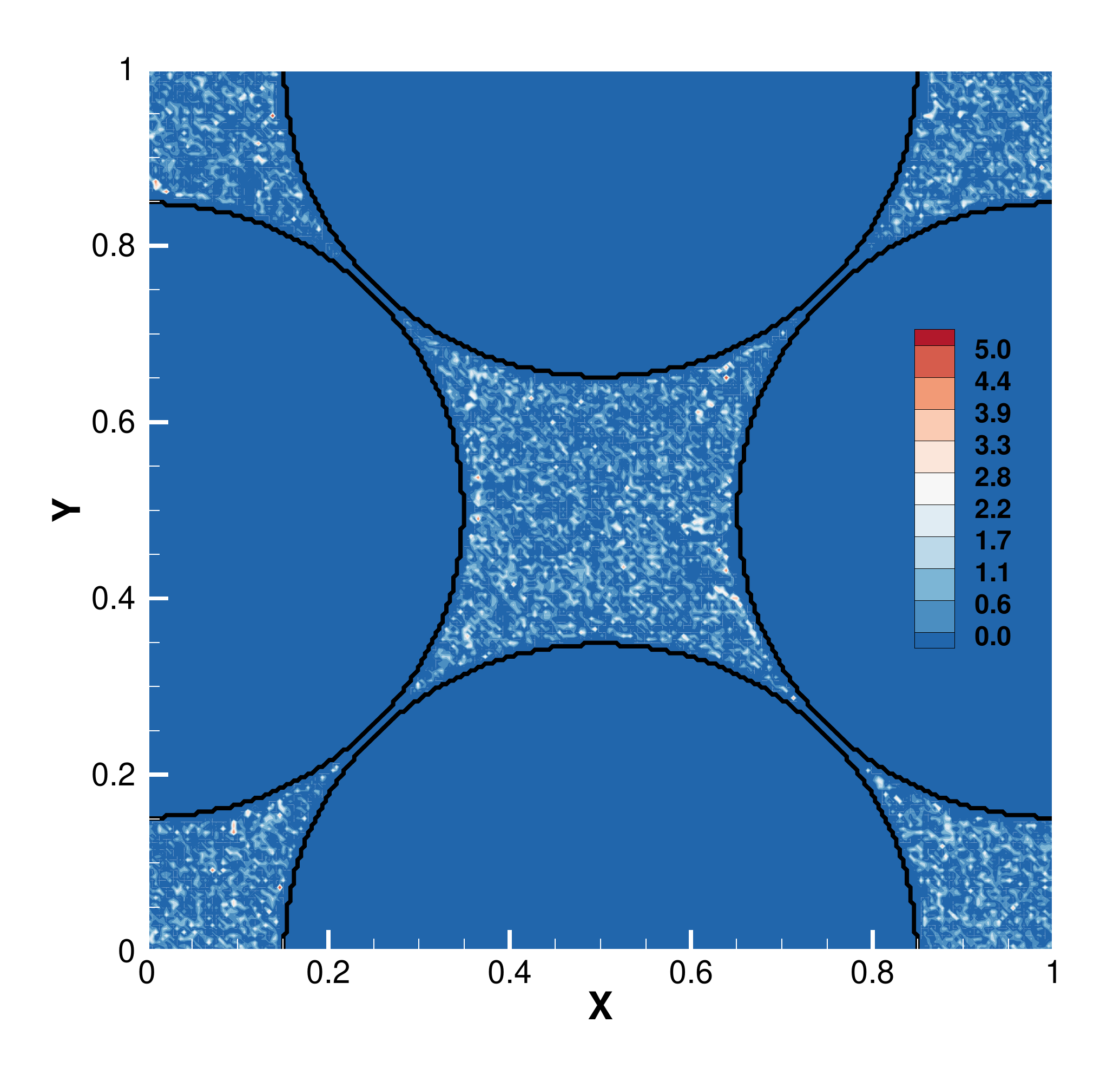}
            \caption{$n^\prime$, $St=0.01$}
             \label{St0.01full}
          \end{subfigure}\\
      \begin{subfigure}[!htpb!]{0.47\textwidth}
         \includegraphics[height=\textwidth]{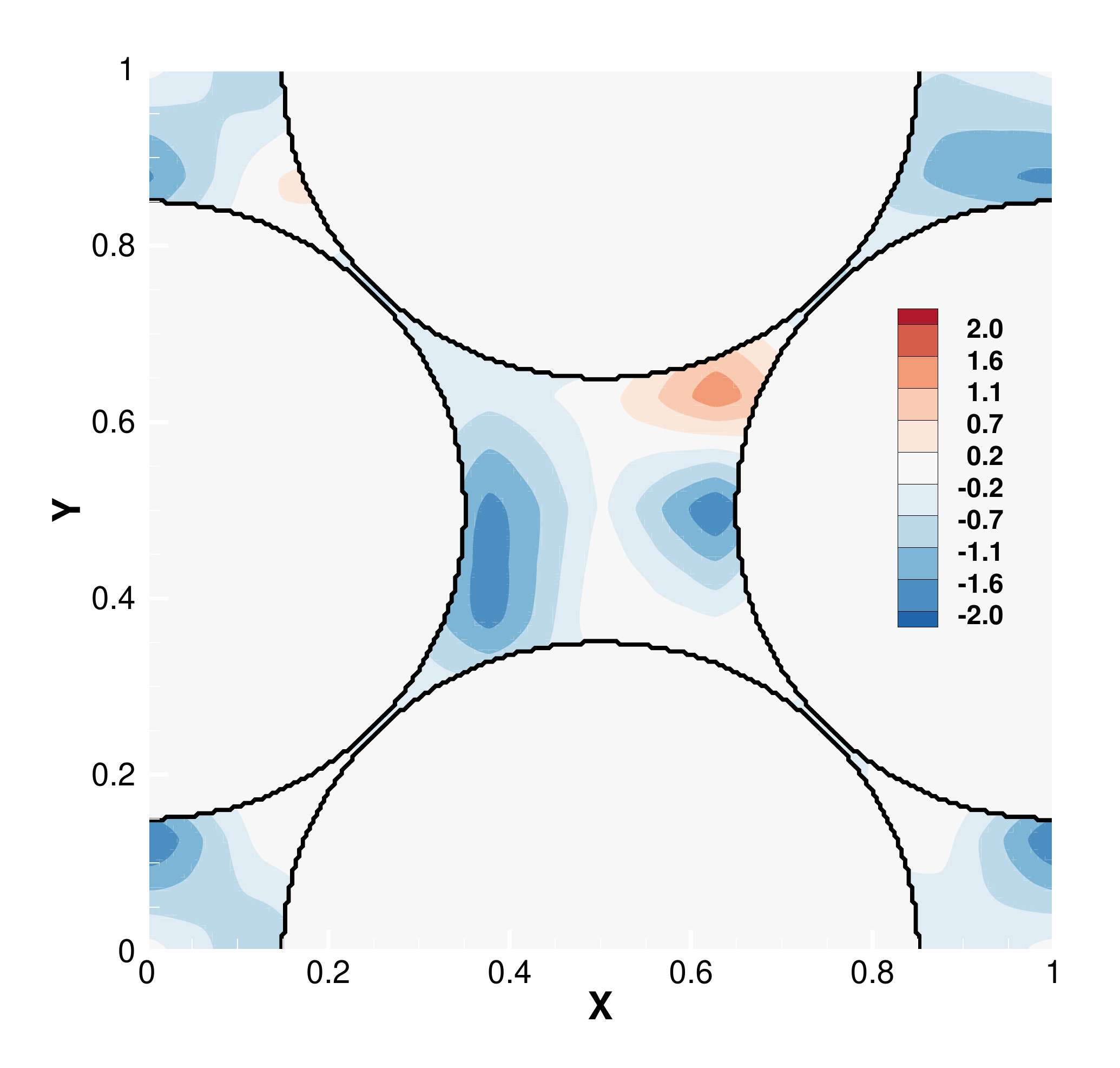}
            \caption{$j=2$, $St=1$}
         \label{St1_J2}
       \end{subfigure}
             \begin{subfigure}[!htpb!]{0.47\textwidth}
         \includegraphics[height=\textwidth]{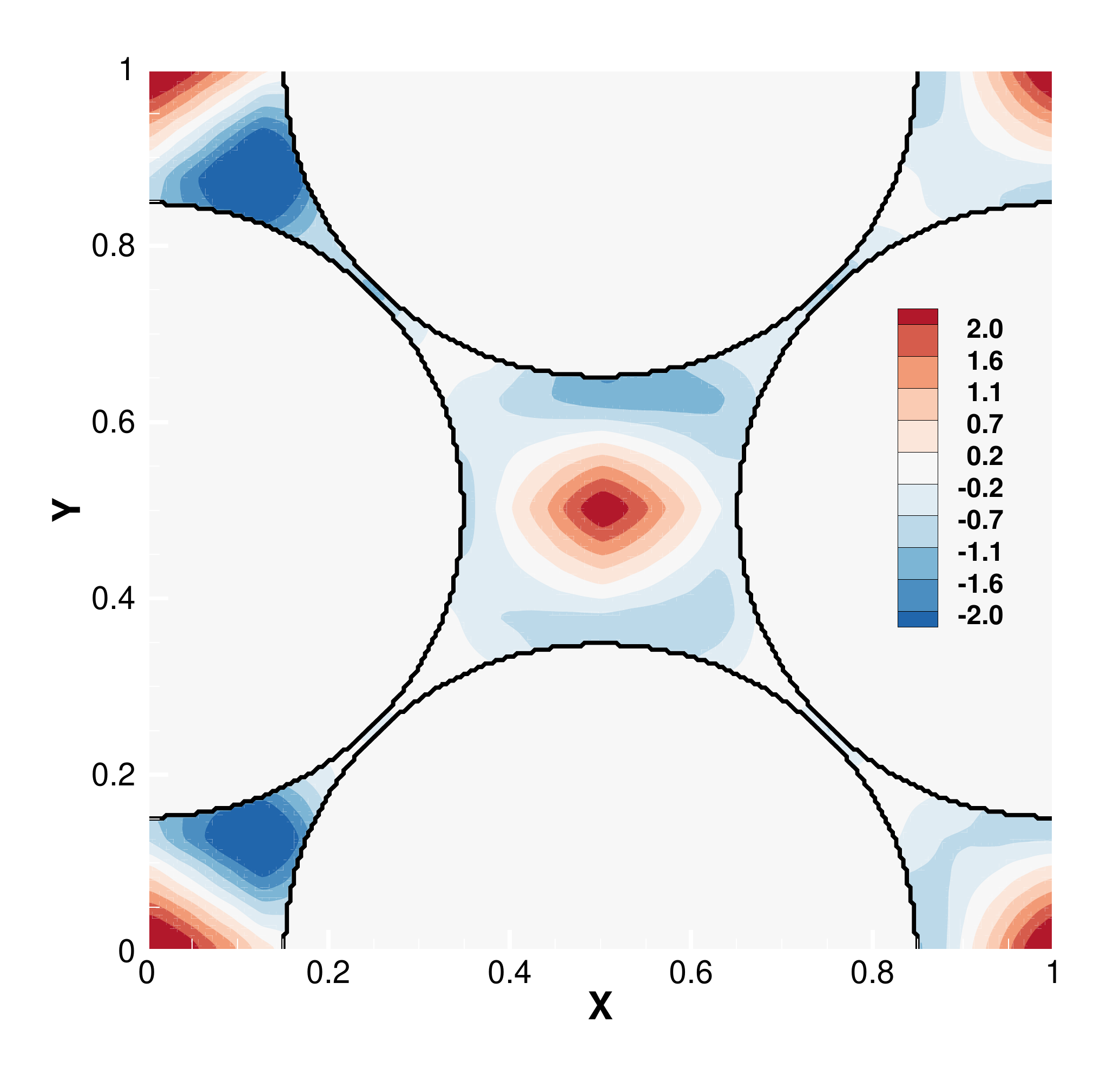}%
        \caption{$j=2$, $St=0.01$}
             \label{St0.01_J2}
       \end{subfigure}\\
             \begin{subfigure}[!htpb!]{0.47\textwidth}
         \includegraphics[height=\textwidth]{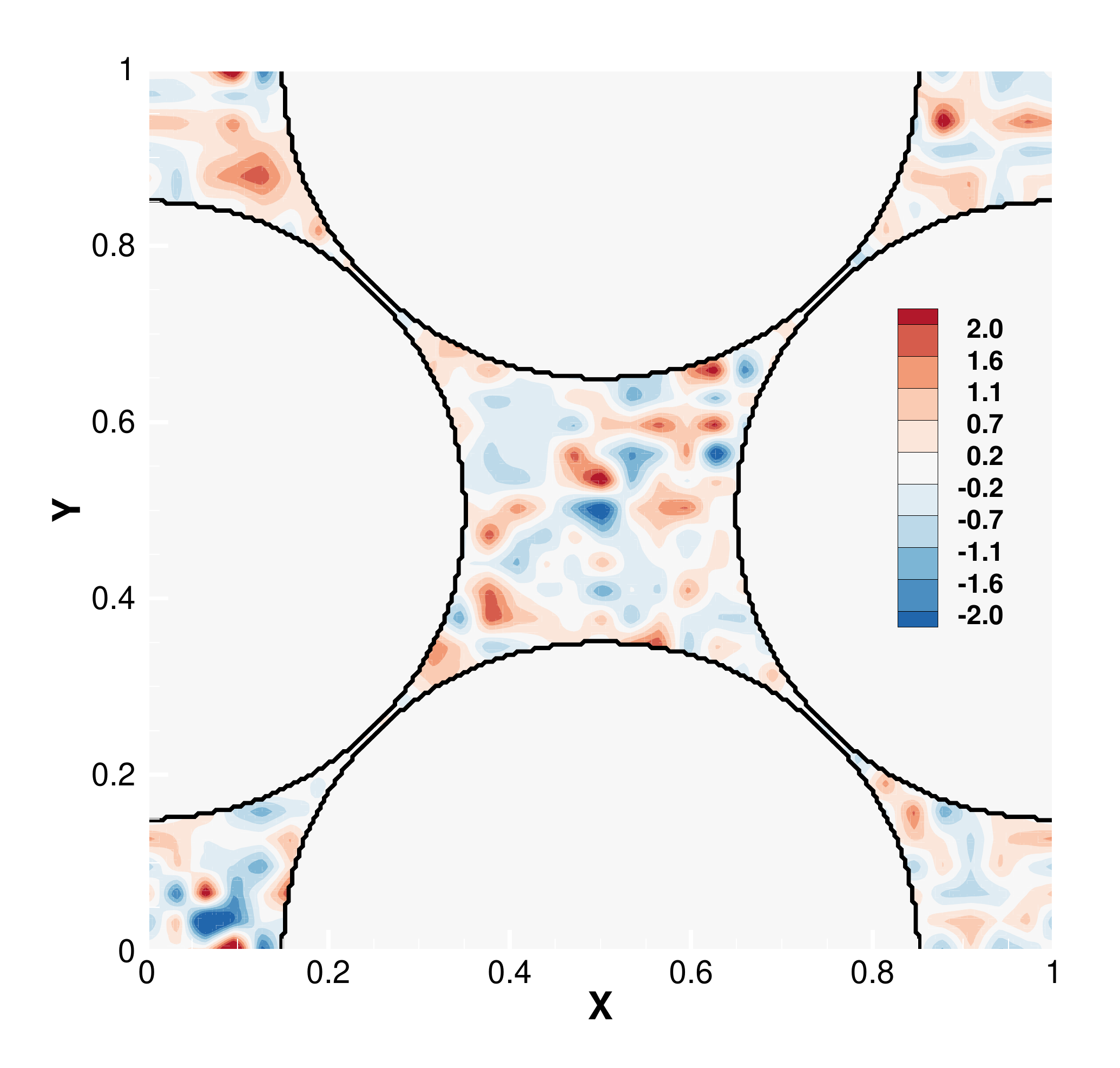}
            \caption{$j=4$, $St=1$}
             \label{St1_J4}
      \end{subfigure}
             \begin{subfigure}[!htpb!]{0.47\textwidth}
         \includegraphics[height=\textwidth]{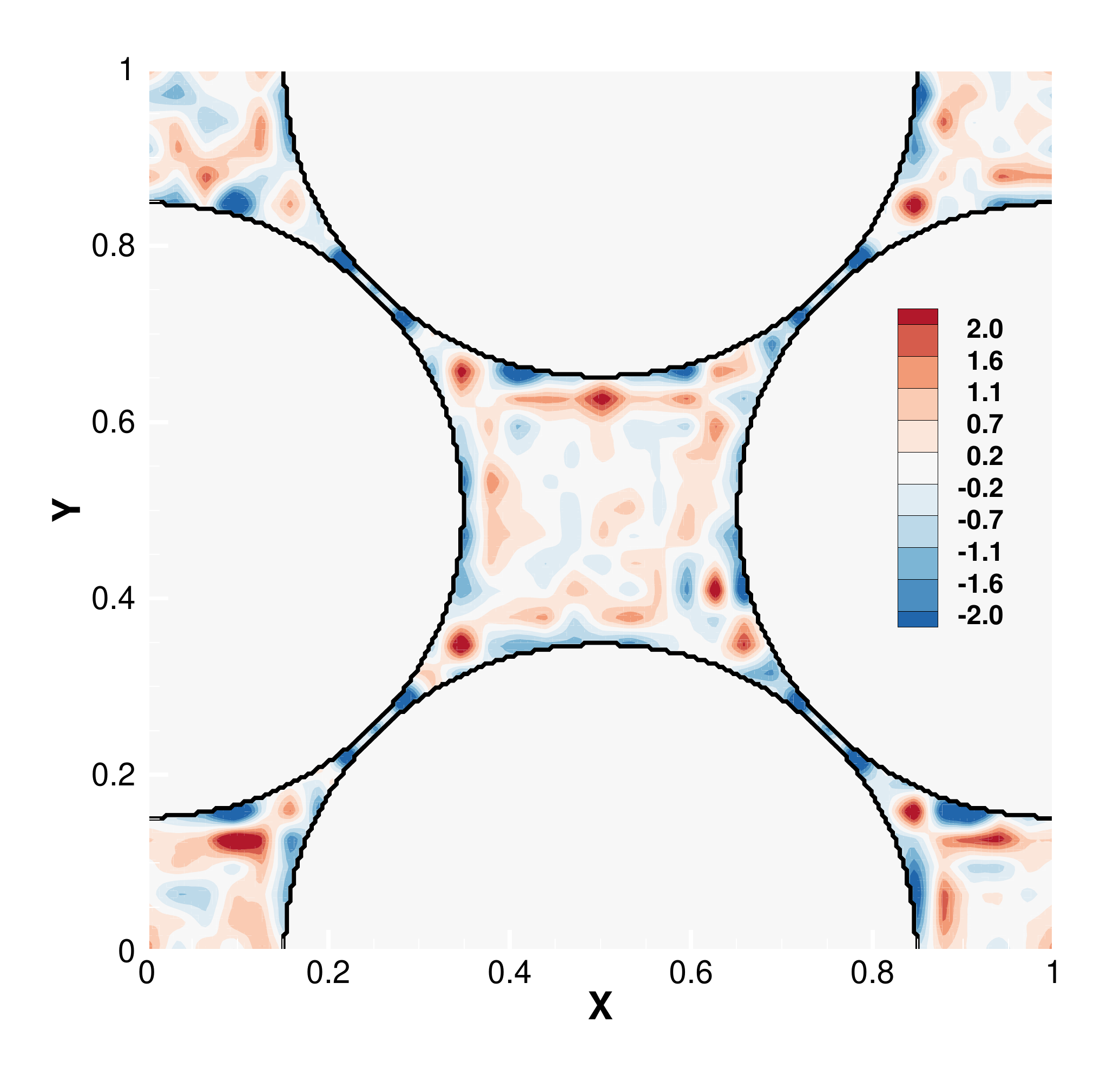}
            \caption{$j=4$, $St=0.01$}
             \label{St0.01_J4}
       \end{subfigure}
       \end{figure}
       
       \begin{figure}\ContinuedFloat
            \centering
             \begin{subfigure}[!htpb!]{0.47\textwidth}\ContinuedFloat
         \includegraphics[height=\textwidth]{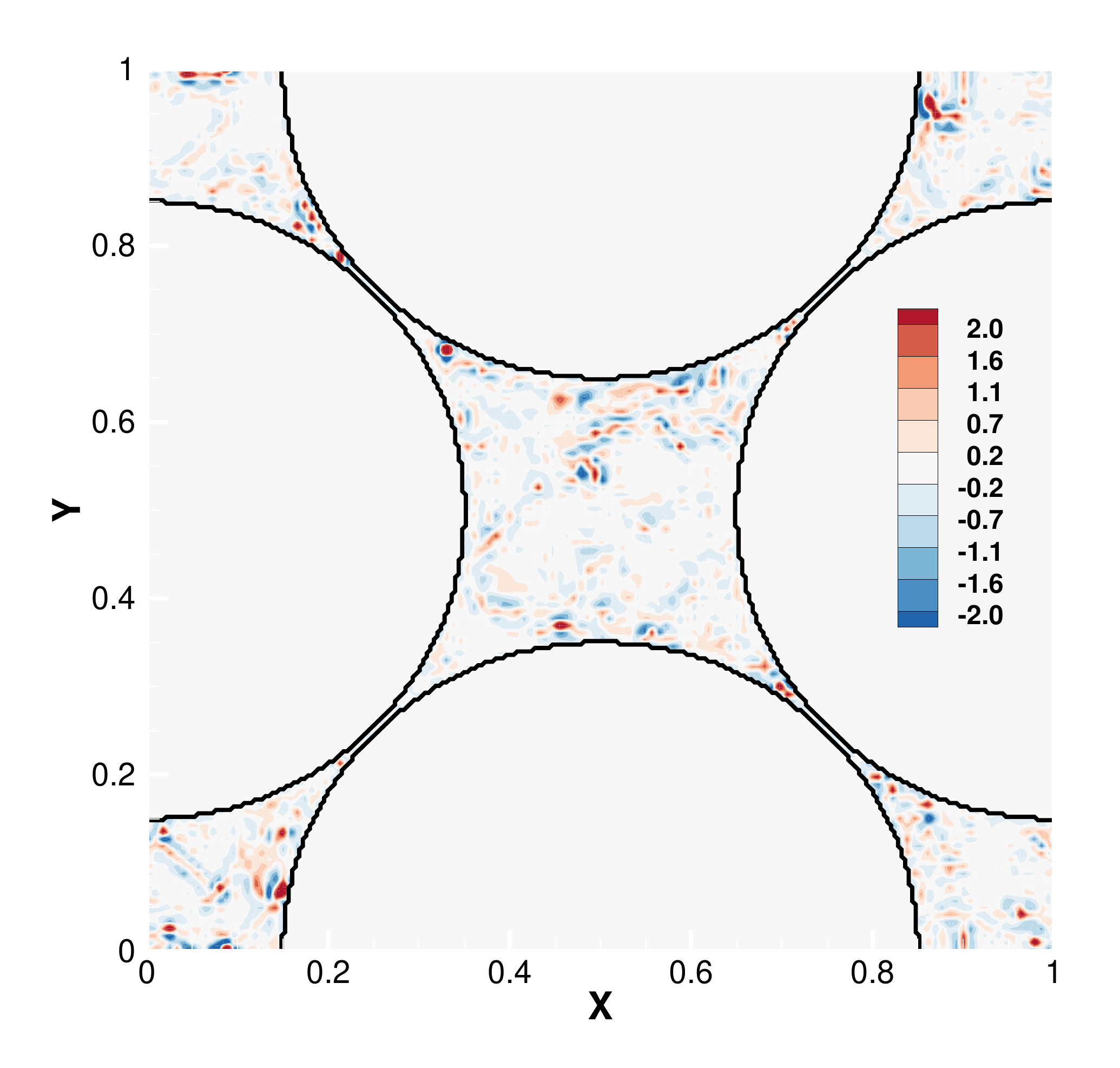}
            \caption{$j=6$, $St=1$}
             \label{St1_J6}
       \end{subfigure}
             \begin{subfigure}[!htpb!]{0.47\textwidth}
         \includegraphics[height=\textwidth]{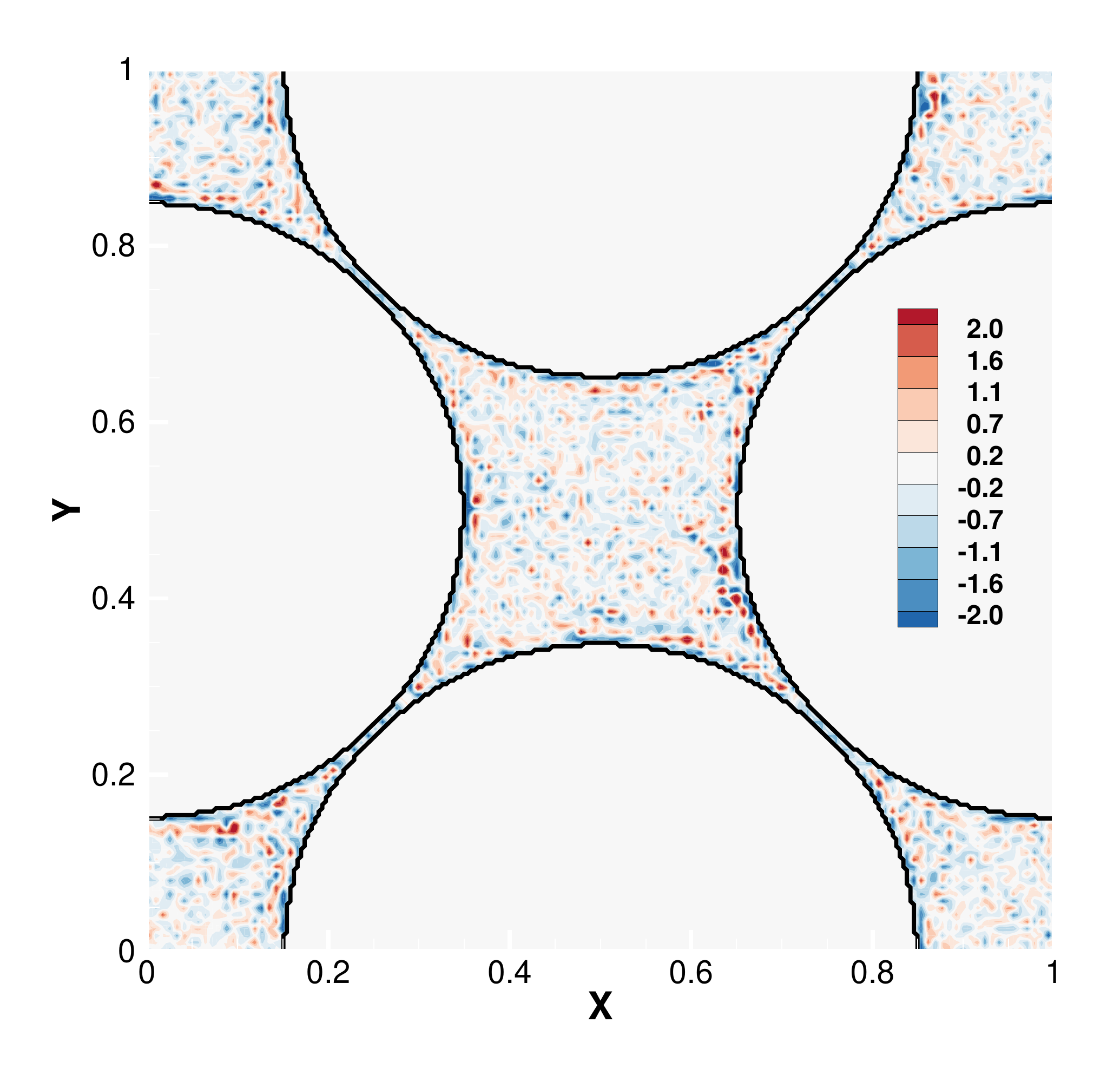}
            \caption{$j=6$, $St=0.01$}
             \label{St0.01_J6}
       \end{subfigure}
	\caption{Instantaneous distribution of normalized particle number density (a,b) and scale contributions 
	($n_j/\sigma[n_j]$) at $j=2$ (c,d), $j=4$ (e,f), and $j=6$ (e,f) in the $xy-$plane for $St=1$ (left) and $St=0.01$ (right).}\label{fig_numden}
\end{figure}

\begin{equation}
n_{j}({\mathbf x}) = \sum_{{ \mu}=1}^7 \sum_{{ i_1=0}}^{2^{ j}-1} \sum_{{i_2=0}}^{2^{j}-1} \sum_{{ i_3=0}}^{2^{ j}-1} {\widetilde n}_{ { j,}   {i_1, i_2, i_3}   }^{ \mu}
{\psi^{\mu}_{j, i_1, i_2, i_3}} ({\mathbf x}),
\end{equation}
with,
\begin{equation}
{\widetilde n}_{ {j}   {i_1 i_2 i_3}   }^{ \mu} = \left< n({\mathbf x}),  {\psi}_{ {j,}   {i_1, i_2, i_3}   }^{ \mu}({\mathbf x}) \right>,
\label{eq:particledensity_scale}
\end{equation}
where $\left<,\right>$ denotes an inner product.
At scale $2^{-j}$, there are $7\times 2^{3j}$ wavelet coefficients for $n({\mathbf x})$. Thus, in total there are $N_g^3$ coefficients for each component of the vector field corresponding to the $N_g^3-1$ wavelet coefficients and the non-vanishing mean value. These coefficients are efficiently computed for $N_g^3-1$ grid points for $n({\mathbf x})$ using fast wavelet transform, which has linear computational complexity. The scale from the wavelet transform and wave number, $k_j$, from the Fourier transform are related as,
\begin{equation}
   k_j = \frac{2\pi}{L} k_{\psi} 2^{-j}, 
\end{equation}
where $k_{\psi} =0.77$ is the centroid wave number of the chosen Coiflet 12 wavelet.

Figure~\ref{fig_numden}a,b shows instantaneous, normalized mean number density contours in the 
$xy-$plane for two Stokes numbers, $St=1$ and $0.01$. Large number density, indicative of highly clustered regions, are clearly observed for $St=1$, but are absent for $St=0.01$. The instantaneous number density field is decomposed using wavelet transform and the scale-dependent fields ($n_j$) normalized by its variance ($\sigma[n_j]=\sqrt{M_2[n_j]}$) are shown for different scales in figure~\ref{fig_numden}c--f. It can be seen that clusters are prominent as scales become smaller (larger $j$), whereas large void regions (negative $n_j$) of size comparable to the pore size are seen at larger scales (smaller $j$). Prominent cluster regions are also seen near the bead surfaces, even for low Stokes number ($St=0.01$). For intermediate scales, both clusters and voids are distributed more intermittently in space. The multiscale nature of clusters and voids is qualitatively clear from these figures. Scale dependent statistics are computed to quantify these differences at various scales.


\subsubsection{Wavelet spectra, scale dependent skewness and flatness}
\label{subsub:waveletresults}
Wavelet-based statistics of the particle number density can be computed as described below. The $q^{th}$ moment of $n_j({\mathbf x})$ is defined as,
\begin{equation}
    M_q[n_j] = \left< (n_j)^q\right>,
\end{equation}
where the mean values $\left< n_j \right>=0$, by giving a central moment and are related to the $q^{th}$ order structure functions~\citep{schneider2004spatial}.

The wavelet energy spectrum of $n_j({\mathbf x})$ can be defined used the second-order moment, $M_2[n_j]$, as
\begin{equation}
E[n_j] = {1 \over \Delta k_j} M_2[n_j],
\end{equation}
where $\Delta k_j = (k_{j+1}-k_j){\rm ln}2$ as given by~\citet{meneveau1991analysis}. 
The energy spectrum obtained from the above equation has the particle number dependence due to the Poisson noise. 
\cite{matsuda2020scale} analytically obtained the effect of the Poisson noise on $M_2[n_j]$ and succeeded in removing the Poisson noise from the wavelet energy spectrum.
The second-order moment for randomly distributed particles in the cubic domain $\Omega = \Omega_f \cup \Omega_s$ is $M_{2,\mathrm{random},\Omega}[n_j] = (7\cdot2^{3j}/N_{p,\Omega})\langle n \rangle_{\Omega}^2$.
For the present case, particles exist only in the fluid domain $\Omega_f$, and $N_p = \phi N_{p,\Omega}$ and $\langle n \rangle = \phi \langle n \rangle_{\Omega}$. 
The energy of the Poisson noise is also reduced by a factor of $\phi$, i.e., $M_{2,\mathrm{random}}[n_j] = \phi M_{2,\mathrm{random},\Omega}[n_j]$, yielding
\begin{equation}
M_{2,{\mathrm{random}}} [ n_j ] = \frac{7\cdot2^{3j}}{N_p} 
\label{eq:M2_random}. 
\end{equation}
Hence, the following definition for the wavelet energy spectrum is used:
\begin{equation}
{E} [n_j]  = \frac{1}{\Delta k_j} 
    \left\{ M_2 [ n_j ] - \frac{7\cdot2^{3j}}{N_p} \right\}
\label{eq:wave_spe},
\end{equation}
where the influence of the Poisson noise has been removed. 
Note that the analytical estimate of Eq.~(\ref{eq:M2_random}) does not contain contribution of the beads geometry. It was found that the energy due to Poisson noise is orders of magnitude smaller for all scales and does not significantly affect the energy spectra of inertial particles. If $M_2[n_j]$ is computed from a realization of random particle distribution in the fluid domain $\Omega_f$, the combined effect of the Poisson noise and geometrical confinement in $\Omega_f$ can be observed.

The asymmetry of the PDF of $n_j({\mathbf x})$ is quantified by the skewness defined as,
\begin{equation}
S[n_j] = {M_3[n_j] \over (M_2[n_j])^{3/2}}.
\end{equation}
The scale-dependent flatness, which measures the intermittency at scale $2^{-j}$, is given as,
\begin{equation}
F[n_j]= {M_4[n_j] \over (M_2[n_j])^{2}},
\end{equation}
and is equal to three at all scales for a Gaussian distribution.
It should be noted that the influence of the Poisson noise on $M_3[n_j]$ and $M_4[n_j]$ cannot be removed by subtracting the moments for randomly distributed particles. 
\cite{matsuda2020scale} introduced the signal-to-noise ratio (SNR) defined as the ratio of the energy spectrum for inertial particles to that for randomly distributed particles, i.e., 
$\mathrm{SNR}=E[n_j]\Delta k_j/M_{2,\mathrm{random}}[n_j]$. 
They confirmed that the effect of the Poisson noise on the statistics is negligibly small when SNR is larger than 10.

Scale dependent wavelet spectra, flatness, and skewness of number density fluctuation are obtained by collecting data over a time span of $15 \, T_{11}^L \approx 30 \, \tau_{\eta}$, where $T_{11}^L$ is the Lagrangian time scale and $\tau_{\eta}$ is the Kolmogorov time scale.  The scale dependent statistics are averaged using about 15 equally spaced snapshots of number density over this time frame.
\begin{figure}
     \centering
    \begin{subfigure}[!htpb!]{0.47\textwidth}
    \includegraphics[height=\textwidth]{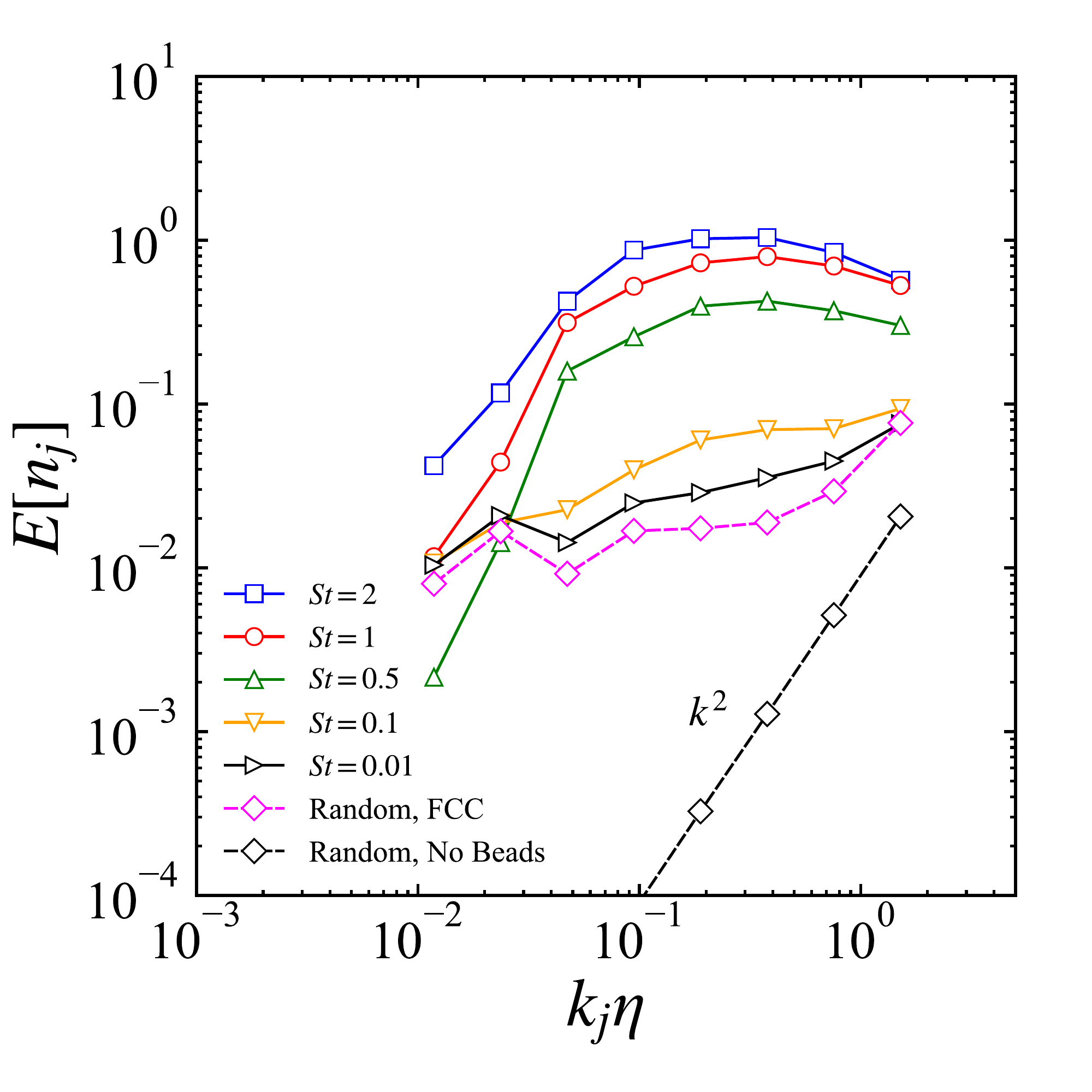}
    \caption{$E\left[n_j\right]$}
    \end{subfigure}
    \begin{subfigure}[!htpb!]{0.47\textwidth}
    \includegraphics[height=\textwidth]{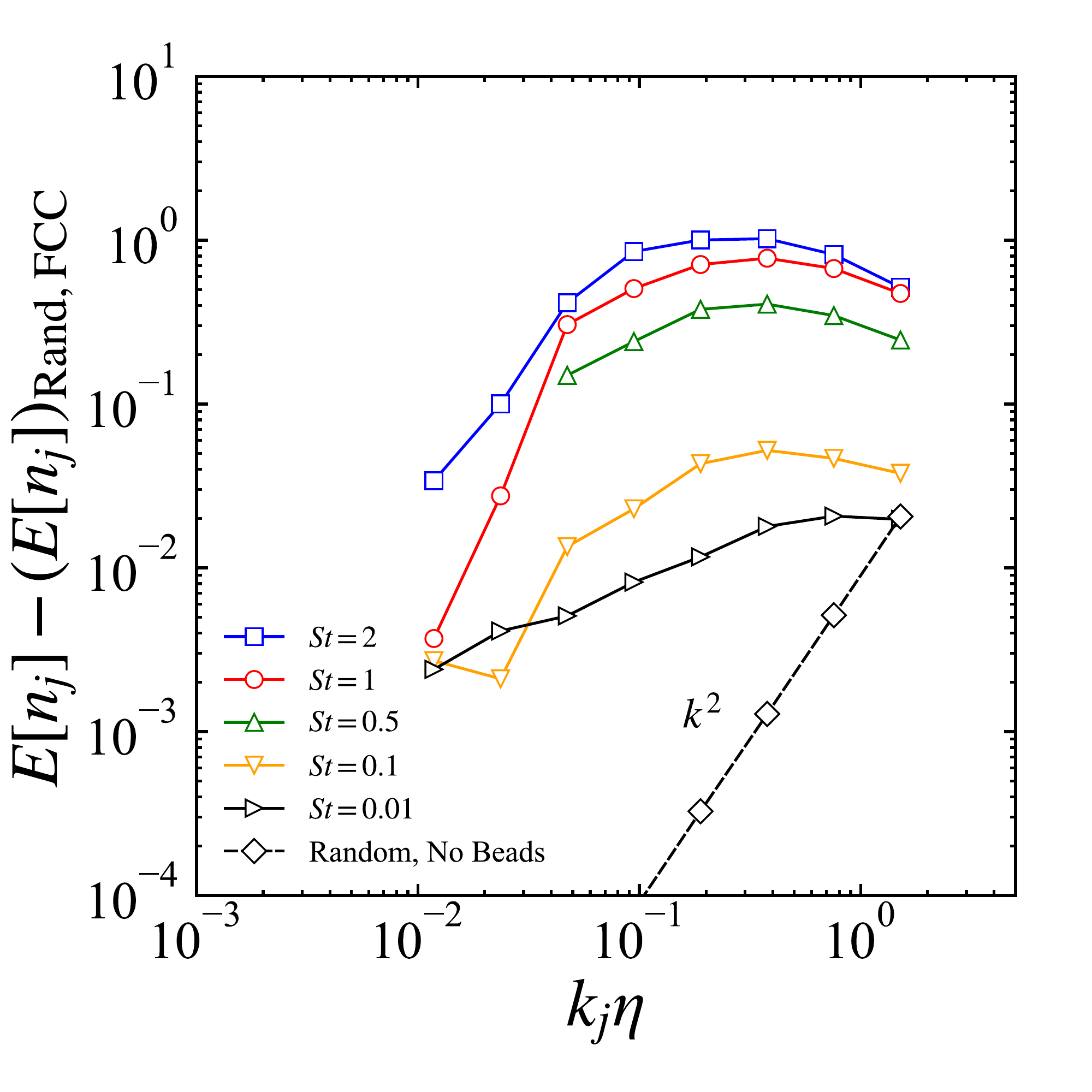}
    \caption{$E\left[n_j\right]-(E\left[n_j\right])_{\rm Random,~FCC}$}
    \end{subfigure}
    \caption{Wavelet energy spectra of particle number density fluctuation $E[n_j]$ for different Stokes numbers. Also shown are energy spectra for Poisson noise ($k^2$ line) in a cubic box (Random, No Beads) and uniform random data in the porous region of the FCC geometry (Random, FCC).}
\label{fig:spectra}
\end{figure}
Figure~\ref{fig:spectra}a presents the wavelet spectra of the number density fluctuations, $E[n_j]$, for different Stokes numbers as a function of the wavenumber, $k_j$, normalized by the Kolmogorov scale, $\eta=(\nu^3/\left<\epsilon\right>)^{1/4}$. Also shown is the spectrum for random particle positions with uniform probability, where the PDF satisfies the Poisson distribution, resulting in $E[n_j]\propto k_j^2$. 
Moreover random particles in the porous region of the FCC geometry are considered for comparison, and the spectrum shows a behavior similar to the $St=0.01$ case. This is expected as random particles correspond to the case $St=0$, representative of fluid tracer particles. Comparing this with random particles in the cubic cell without the beads, i.e. the $k^2$ scaling, quantifies the influence of the geometrical confinement. At small scales increasing energy is observed and a similar magnitude as for $St \le 0.1$.
For $St=$ 0.5, 1, and 2, it is seen that the spectrum first increases with a increasing $k_j \eta$ (large scales) and then gradually decreases for large $k_j\eta$ (small scales). The peak in the spectrum is found to gradually shift towards larger scales (smaller $k_j \eta$). This observation is consistent with the clustering of inertial particles in homogeneous, isotropic turbulence~\citep{matsuda2020scale}. In addition, for higher Stokes numbers, the energy $E[n_j]$ is generally higher for all $k_j \eta$. For $St<0.5$, the peak observed in spectrum at higher $St$, is not seen and $E[n_j]$ increases monotonically with $k_j \eta$. 
Subtracting the noise in the porous region, shown in Figure~\ref{fig:spectra}b, allows to recover the peaks for small $St$. The peak values become higher for similar $k_j \eta$, which indicates that the void scale is almost constant. This is consistent with results for homogeneous, isotropic turbulence~\citep{matsuda2014influence}.
 For isotropic turbulence, with low Stokes numbers, the spectra show a steeper slope close to $k^{-1}$ at small scales ($k_j \eta \gtrsim 0.4$). This difference in slopes at low $St$ in the presence of beads is attributed to the effect of inertial particle collisions with the bead surfaces as well as the geometric confinement effect of the flow.
 
\begin{figure}
     \centering
       \begin{subfigure}[!htpb!]{0.47\textwidth}
             \includegraphics[height=\textwidth]{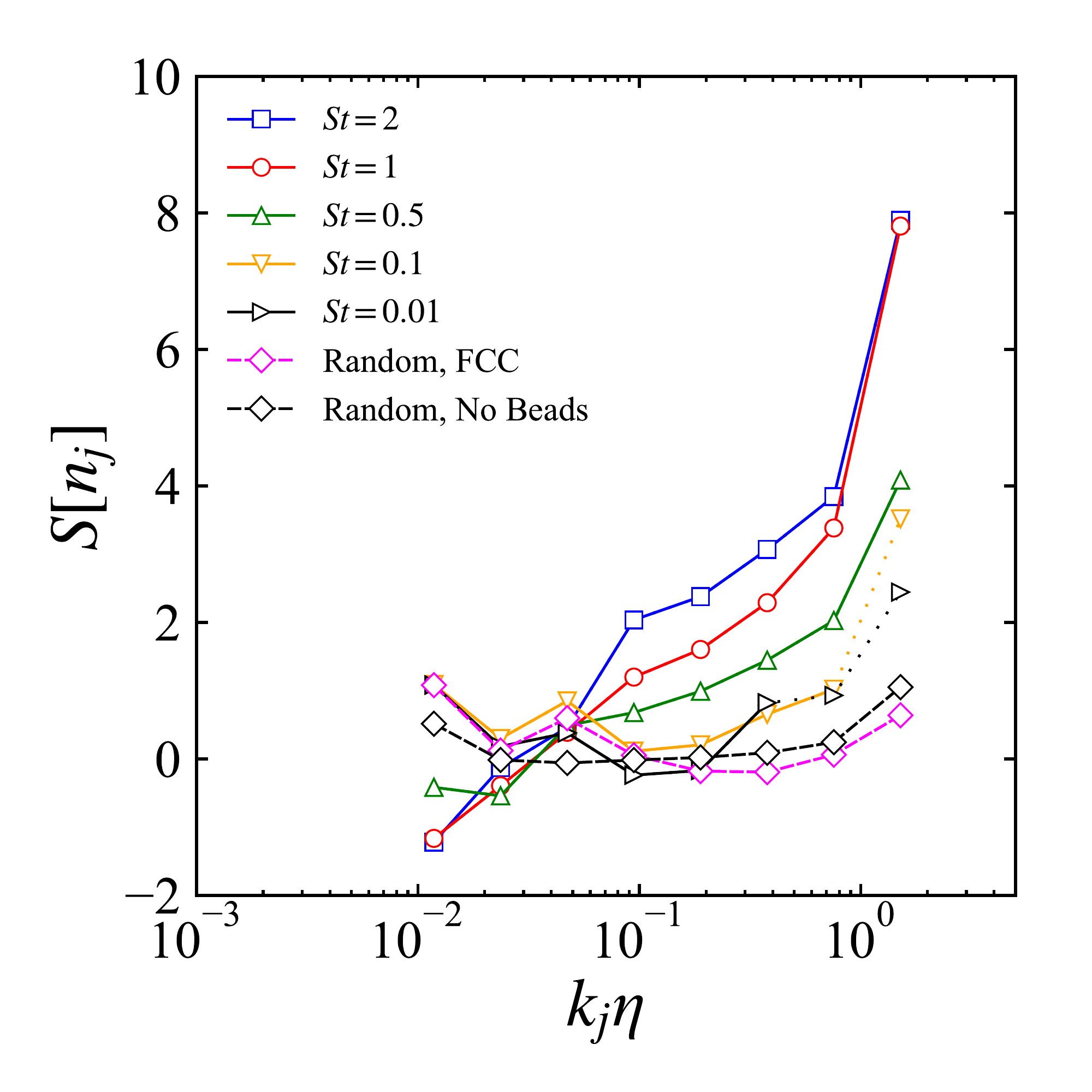} 
             \caption{Skewness}
         \label{skewness}
        \end{subfigure}
        \begin{subfigure}[!htpb!]{0.47\textwidth}
            \includegraphics[height=\textwidth]{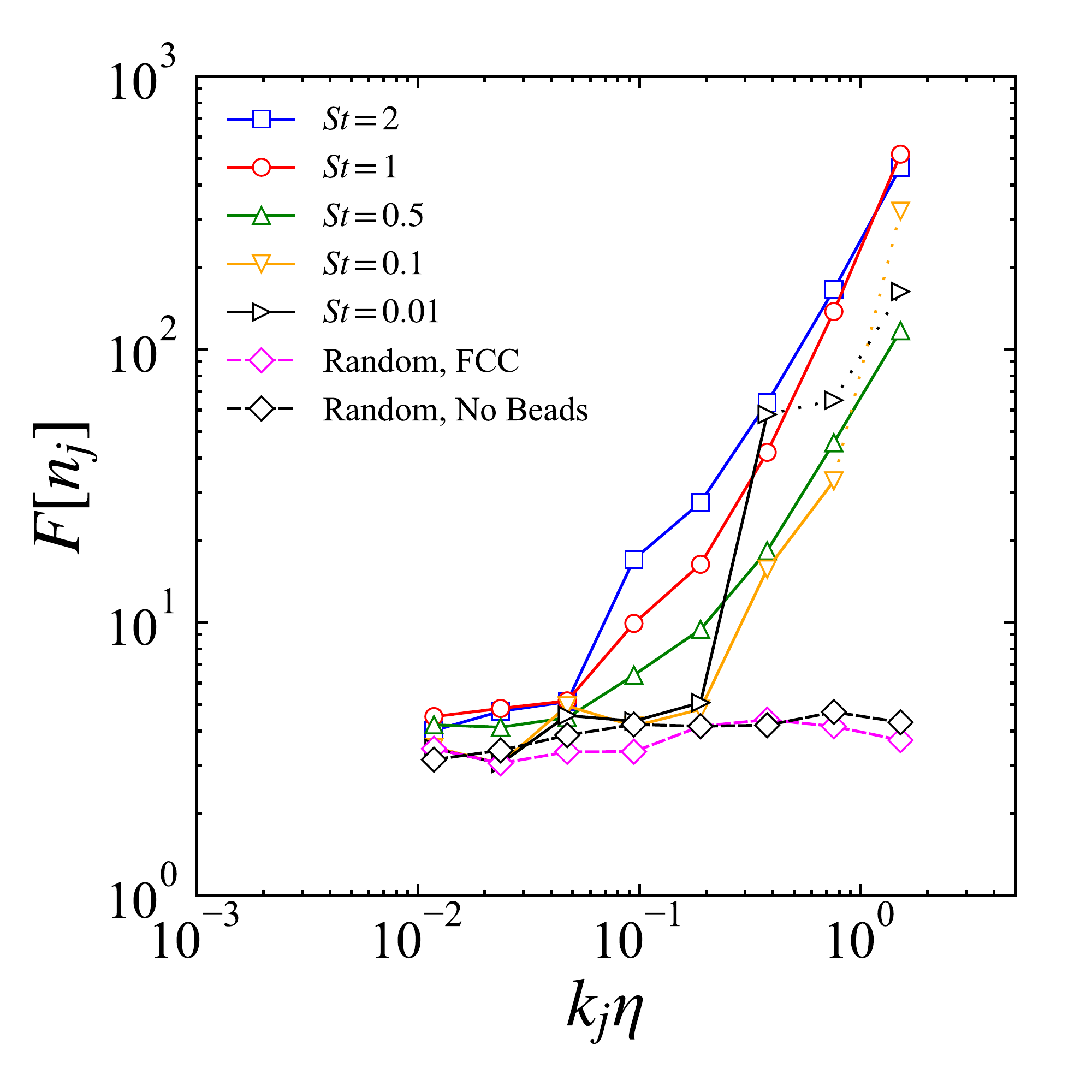}
            \caption{Flatness}
             \label{flatness}
          \end{subfigure}
	\caption{Scale-dependent  skewness $S[n_j]$ (left) and flatness $F[n_j]$ (right) for different Stokes numbers and random particles in a cubic box (Random, No Beads) and uniform random data in the porous region of the FCC geometry (Random, FCC).
	Solid lines connect the symbols at scales for which $\mathrm{SNR} \ge 10$, and dotted lines are used otherwise.} 
	\label{FlatnessSkewness}
\end{figure}

Flatness and skewness statistics of number density fluctuations are computed for various Stokes numbers, as well as randomly distributed particles for the cubic box and the porous region, and are shown in figure~\ref{FlatnessSkewness}a,b. The statistics of randomly distributed particles correspond to those of fluid tracer particles ($St=0$). The number density of the
fluid particles is uniform due to the volume preserving nature of the incompressible flow and particle clustering is absent resulting in zero skewness and flatness close to 3 as expected. 
For random data in in the porous region of the FCC geometry  similar results were obtained with only minor differences.
For all inertial particles with different Stokes numbers, the flatness values increase with decreasing scales, showing that intermittency of particle clustering is high at smaller scales. This is also seen qualitatively in the contour plots (figure~\ref{fig_numden}e--h) of number density fluctuations at smaller scales. For $St>0.5$, the flatness values for each scale ($k_j \eta$) decreases with decreasing Stokes number. However, for small Stokes numbers, there is an increase in intermittency at intermediate and smaller scales ($k_j \eta > 0.2$) compared to larger Stokes numbers. Note that for lower Stokes numbers, particle clustering is typically not significant (figure~\ref{fig_numden}b). Thus, this increase in flatness at smaller scales may be a result of collision of particles with bead surfaces. 

For inertial particles at all Stokes numbers, skewness values also increase with decreasing scales. As shown by~\citet{matsuda2020scale}, positive skewness of number density fluctuations indicate high probablity of large positive values of $n_j({\mathbf x})$, that is prominent clusters, whereas negative skewness corresponds to large excursions of negative values of $n_j({\mathbf x})$, that is  void-pronounced structures.
Increase in positive skewness at smaller scales for all Stokes numbers implies more prominence of clusters. At $St=0.01$ \Keigo{and} $0.1$, the skewness remains close to zero at intermediate and large scales, similar to the random particle statistics. However, at smaller scales, there is an increase in positive skewness for these Stokes numbers. This again is conjectured to be clustering of these particles owing to collisions with the bead surfaces. Finally, small negative values of skewness observed for $St>0.5$ at large scales are indicative of large regions of voids. This is also qualitatively confirmed in the visualizations shown in figure~\ref{fig_numden}.

\section{Summary and conclusion}\label{summ}
 Fully resolved direct numerical simulation of a turbulent flow through the confined geometry of a triply periodic, face-centered cubic (FCC) porous unit cell was performed using a Cartesian grid-based fictitious domain method. The low porosity of the flow geometry gives rise  to  rapid  acceleration  and deceleration of the mean flow with presence of three-dimensional helical motions, weak wake-like structures behind the bead spheres, stagnation and jet-impingement-like flows together with merging and spreading jets in the main pore. Details of turbulence characteristics in this confined geometry of the porous cell for a range of Reynolds numbers spanning unsteady inertial, transitional, and turbulent flow were characterized in detail in prior work~\citep{he2019characteristics}. 
In this work, emphasis is placed on clustering dynamics of inertial particles in a turbulent flow through the porous cell. Specifically, how particle-wall interactions affect clustering and deposition mechanisms at different particle Stokes number were studied. To this end, point particles were advanced for four different Stokes numbers $St$ using one-way coupling and assuming perfectly elastic wall collisions for a single pore Reynolds number of $Re_H = 500$, corresponding to inertial, turbulent flow. About ten million particles for each Stokes number were introduced into the flow and tracked over several flow through times to provide meaningful statistics on inertial particle dynamics of clustering, void formation, and transport inside the confined geometry of the porous medium.

    Tools for studying inertial particle dynamics and clustering, previously developed for homogeneous flows, have been adapted being taken into account the curved flow geometry of the bead walls in the porous cell. Mirror particles were used in the Voronoi analysis and adjusted for tesselation in presence of bead walls.
%
	The probability distribution of the Voronoi volumes for the different Stokes numbers quantified the departure from the gamma distribution and allowed to assess the influence of both the geometry and the flow, in comparison to isotropic turbulence. It was found that the geometry only impacts the small volumes below $10^{-2}$ to the mean Voronoi volume.
	 Cluster formation and destruction was quantified by analyzing the time change of the Voronoi volumes for the different Stokes numbers. This Lagrangian approach yields a time discrete measure of the spatial divergence of the particle velocity. It was found that the PDFs of the divergence, which are symmetric, are becoming wider for increasing Stokes number, as the variance increases. In contrast the tails of the PDFs are becoming shallower with Stokes number.
	 The conditional average of the divergence as a function of volume is found to be positive for small volumes and negative for large volumes. This explains that cluster formation is present for large volumes, while cluster destruction is more prominent for small volumes. Moreover, these effects are amplified with the Stokes number.
	 These findings are similar to what has been found for isotropic turbulence \citep{oujia2020jfm}.
	%
%
    Wavelet-based multiscale statistical analyses were  applied to particle number density fields in the flow through the porous geometry.
	By decomposing the number density fields into orthogonal wavelets, scale-dependent statistics of number density distribution were computed. The wavelet energy spectra showed that the peak of clustering gradually shifts towards larger scale as the Stokes number increases. 
	To reduce the influence of bead geometry, the difference of each energy spectrum for inertial particles from that for randomly distributed particles only in fluid domain was also computed. This allowed to identify the peaks in the spectra for $St \le 0.1$.
    Scale-dependent skewness and flatness of the particle density quantified the intermittent void and cluster distribution statistically. The positive skewness values at smaller scales found for all Stokes numbers confirm the observed small scale prominent clusters. Negative skewness values for $St >0.5$ quantify the presence of prominent void regions. The flatness values which are increasing with decreasing scale and that the values become even larger for larger Stokes number confirm the strongly intermittent cluster distribution.

\medskip

In conclusion both static and dynamic analyses of particle clustering in a porous cell have been performed.
With scale-dependent analyses of snapshots of particle density distributions (static analyses), voids and clusters were quantified statistically. Focusing on intermittency, a signature of void and clusters was observed in higher order statistics. 
With the Lagrangian analysis of Voronoi tesselations (dynamic analysis), the convergence and divergence of particle velocity were computed providing thus an explanation for cluster formation and destruction. 
A comparison with results for homogeneous isotropic turbulence, showed many similarities and also pointed out differences due to the flow geometry, in particular for small volumes.
Combining the multiscale statistical analyses with the Lagrangian formulation of the Voronoi tesselation constitutes an interesting perspective of this work for quantifying scale-dependent divergence and convergence of the particle velocity and the related inter-scale transfer. The analysis presented can be extended in the future to vorticity fields to obtain three-dimensional directional information at different scales.


\medskip

\begin{acknowledgments}
This work was initiated during SVA's visiting scientist/Professor position at Aix-Marseille University. Funding for this position was provided by CNRS, France and Aix-Marseille University. SVA acknowledges kind hospitality during his visit and stay in Marseille. Funding from NSF award\#2053248 is also gratefully acknowledged. Simulations were performed at the Texas Advanced Computing Center's (TACC) Stampede2 and Frontera systems.  BK, TO and KS thankfully acknowledge Centre de Calcul Intensif d’Aix-Marseille for granting access to its high performance computing resources.
TO and KS thankfully acknowledge financial support from Agence Nationale de la Recherche, project ANR-20-CE46-0010-01.
XH thanks Dr. Timothy Scheibe from Pacific Northwest National Laboratory (PNNL) for supporting funding from the DOE Office of Biological and Environmental Research, Subsurface Biogeochemical Research program, through the PNNL Subsurface Science Scientific Focus Area project (http://sbrsfa.pnnl.gov/).
\end{acknowledgments}

\bibliographystyle{jfm}
\bibliography{Porous}

\providecommand{\noopsort}[1]{}\providecommand{\singleletter}[1]{#1}%
\begin{thebibliography}{48}
\expandafter\ifx\csname natexlab\endcsname\relax\def\natexlab#1{#1}\fi
\def\au#1{#1} \def\ed#1{#1} \def\yr#1{#1}\def\at#1{#1}\def\jt#1{\textit{#1}}
  \def\bt#1{#1}\def\bvol#1{\textbf{#1}} \def\vol#1{#1} \def\pg#1{#1}
  \def\publ#1{#1}\def\arxiv#1{#1}\def\org#1{#1}\def\st#1{\textit{#1}}

\bibitem[Agbangla {\em et~al.\/}(2012)Agbangla, Climent \&
  Bacchin]{agbangla2012experimental}
{\sc \au{Agbangla, Gbedo~Constant}, \au{Climent, {\'E}ric} \& \au{Bacchin,
  Patrice}} \yr{2012}  \at{Experimental investigation of pore clogging by
  microparticles: Evidence for a critical flux density of particle yielding
  arches and deposits}.  \jt{Separation and purification technology}
  \bvol{101},  \pg{42--48}.

\bibitem[Apte {\em et~al.\/}(2009)Apte, Martin \& Patankar]{apte2009frs}
{\sc \au{Apte, S.~V.}, \au{Martin, M.} \& \au{Patankar, N.~A.}} \yr{2009}
  \at{A numerical method for fully resolved simulation ({FRS}) of rigid
  particle-flow interactions in complex flows}.  \jt{Journal of Computational
  Physics}  \bvol{228}~(8),  \pg{2712--2738}.

\bibitem[Aris(1999)]{aris1999elementary}
{\sc \au{Aris, Rutherford}} \yr{1999} {\em Elementary chemical reactor
  analysis\/}.  \publ{Courier Corporation}.

\bibitem[Aurenhammer(1991)]{aurenhammer1991voronoi}
{\sc \au{Aurenhammer, Franz}} \yr{1991}  \at{Voronoi diagrams—a survey of a
  fundamental geometric data structure}.  \jt{ACM Computing Surveys (CSUR)}
  \bvol{23}~(3),  \pg{345--405}.

\bibitem[Barber {\em et~al.\/}(1996)Barber, Dobkin \&
  Huhdanpaa]{barber1996quickhull}
{\sc \au{Barber, C~Bradford}, \au{Dobkin, David~P} \& \au{Huhdanpaa, Hannu}}
  \yr{1996}  \at{The quickhull algorithm for convex hulls}.  \jt{ACM
  Transactions on Mathematical Software (TOMS)}  \bvol{22}~(4),  \pg{469--483}.

\bibitem[Bassenne {\em et~al.\/}(2017)Bassenne, Urzay, Schneider \&
  Moin]{bassenne2017extraction}
{\sc \au{Bassenne, Maxime}, \au{Urzay, Javier}, \au{Schneider, Kai} \&
  \au{Moin, Parviz}} \yr{2017}  \at{Extraction of coherent clusters and grid
  adaptation in particle-laden turbulence using wavelet filters}.  \jt{Physical
  Review Fluids}  \bvol{2}~(5),  \pg{054301}.

\bibitem[Bec {\em et~al.\/}(2007)Bec, Biferale, Cencini, Lanotte, Musacchio \&
  Toschi]{bec2007heavy}
{\sc \au{Bec, Jeremie}, \au{Biferale, Luca}, \au{Cencini, Massimo},
  \au{Lanotte, Alessandra}, \au{Musacchio, Stefano} \& \au{Toschi, Federico}}
  \yr{2007}  \at{Heavy particle concentration in turbulence at dissipative and
  inertial scales}.  \jt{Physical Review Letters}  \bvol{98}~(8),  \pg{084502}.

\bibitem[Carlson {\em et~al.\/}(1992)Carlson, Gurley, King, Price-Smith \&
  Waters]{carlson1992sand}
{\sc \au{Carlson, Jon}, \au{Gurley, Derrel}, \au{King, George},
  \au{Price-Smith, Colin} \& \au{Waters, Frank}} \yr{1992}  \at{Sand control:
  Why and how?}  \jt{Oilfield Review}  \bvol{4}~(4),  \pg{41--53}.

\bibitem[Coleman \& Vassilicos(2009)]{coleman2009unified}
{\sc \au{Coleman, S.W.} \& \au{Vassilicos, J.C.}} \yr{2009}  \at{A unified
  sweep-stick mechanism to explain particle clustering in two-and
  three-dimensional homogeneous, isotropic turbulence}.  \jt{Physics of Fluids}
   \bvol{21}~(11),  \pg{113301}.

\bibitem[Cook {\em et~al.\/}(2004)Cook, Lee, DiGiovanni, Bronowski, Perkins \&
  Williams]{cook2004discrete}
{\sc \au{Cook, BK}, \au{Lee, MY}, \au{DiGiovanni, AA}, \au{Bronowski, DR},
  \au{Perkins, ED} \& \au{Williams, JR}} \yr{2004}  \at{Discrete element
  modeling applied to laboratory simulation of near-wellbore mechanics}.
  \jt{International Journal of Geomechanics}  \bvol{4}~(1),  \pg{19--27}.

\bibitem[Dai \& Grace(2010)]{dai2010blockage}
{\sc \au{Dai, Jianjun} \& \au{Grace, John~R}} \yr{2010}  \at{Blockage of
  constrictions by particles in fluid--solid transport}.  \jt{International
  Journal of Multiphase Flow}  \bvol{36}~(1),  \pg{78--87}.

\bibitem[Daubechies(1993)]{daubechies1993ten}
{\sc \au{Daubechies, Ingrid}} \yr{1993} {\em Ten lectures on wavelets\/}.
  \publ{Society of Industrial and Applied Mathematics}.

\bibitem[Dixon \& Nijemeisland(2001)]{Dixon2001aa}
{\sc \au{Dixon, A.~G.} \& \au{Nijemeisland, M.}} \yr{2001}  \at{{CFD} as a
  design tool for fixed-bed reactors}.  \jt{Industrial \& Engineering Chemistry
  Research}  \bvol{40}~(23),  \pg{5246--5254}.

\bibitem[Eaton \& Fessler(1994)]{eaton1994preferential}
{\sc \au{Eaton, John~K} \& \au{Fessler, JR}} \yr{1994}  \at{Preferential
  concentration of particles by turbulence}.  \jt{International Journal of
  Multiphase Flow}  \bvol{20},  \pg{169--209}.

\bibitem[Esmaily-Moghadam \& Mani(2016)]{esmaily2016analysis}
{\sc \au{Esmaily-Moghadam, Mahdi} \& \au{Mani, Ali}} \yr{2016}  \at{Analysis of
  the clustering of inertial particles in turbulent flows}.  \jt{Physical
  Review Fluids}  \bvol{1}~(8),  \pg{084202}.

\bibitem[Farge(1992)]{farge1992}
{\sc \au{Farge, M.}} \yr{1992}  \at{Wavelet transforms and their applications
  to turbulence}.  \jt{Annual Review of Fluid Mechanics}  \bvol{24},
  \pg{395--457}.

\bibitem[Farge \& Schneider(2015)]{farge2015}
{\sc \au{Farge, M.} \& \au{Schneider, K.}} \yr{2015}  \at{Wavelet transforms
  and their applications to {M}{H}{D} and plasma turbulence: a review}.
  \jt{Journal of Plasma Physics}  \bvol{81}~(6),  \pg{435810602}.

\bibitem[Ferenc \& N{\'e}da(2007)]{ferenc2007size}
{\sc \au{Ferenc, J{\'a}rai-Szab{\'o}} \& \au{N{\'e}da, Zolt{\'a}n}} \yr{2007}
  \at{On the size distribution of poisson voronoi cells}.  \jt{Physica A:
  Statistical Mechanics and its Applications}  \bvol{385}~(2),  \pg{518--526}.

\bibitem[Finn \& Apte(2013)]{justin2013rela}
{\sc \au{Finn, J.} \& \au{Apte, S.~V.}} \yr{2013}  \at{Relative performance of
  body fitted and fictitious domain simulations of flow through fixed packed
  beds of spheres}.  \jt{International Journal of Multiphase Flow}  \bvol{56},
  \pg{54--71}.

\bibitem[Goto \& Vassilicos(2006)]{goto2006self}
{\sc \au{Goto, Susumu} \& \au{Vassilicos, J.C.}} \yr{2006}  \at{Self-similar
  clustering of inertial particles and zero-acceleration points in fully
  developed two-dimensional turbulence}.  \jt{Physics of Fluids}
  \bvol{18}~(11),  \pg{115103}.

\bibitem[He {\em et~al.\/}(2018)He, Apte, Schneider \& Kadoch]{he2018angular}
{\sc \au{He, Xiaoliang}, \au{Apte, Sourabh}, \au{Schneider, Kai} \& \au{Kadoch,
  Benjamin}} \yr{2018}  \at{Angular multiscale statistics of turbulence in a
  porous bed}.  \jt{Physical Review Fluids}  \bvol{3}~(8),  \pg{084501}.

\bibitem[He {\em et~al.\/}(2019)He, Apte, Finn \& Wood]{he2019characteristics}
{\sc \au{He, Xiaoliang}, \au{Apte, Sourabh~V}, \au{Finn, Justin~R} \& \au{Wood,
  Brian~D}} \yr{2019}  \at{Characteristics of turbulence in a face-centred
  cubic porous unit cell}.  \jt{Journal of Fluid Mechanics}  \bvol{873},
  \pg{608--645}.

\bibitem[Hester {\em et~al.\/}(2017)Hester, Cardenas, Haggerty \&
  Apte]{hester2017importance}
{\sc \au{Hester, Erich~T}, \au{Cardenas, M~Bayani}, \au{Haggerty, Roy} \&
  \au{Apte, Sourabh~V}} \yr{2017}  \at{The importance and challenge of
  hyporheic mixing}.  \jt{Water Resources Research}  \bvol{53}~(5),
  \pg{3565--3575}.

\bibitem[Hill \& Koch(2002)]{hill2002trans}
{\sc \au{Hill, R.~J.} \& \au{Koch, D.~L.}} \yr{2002}  \at{The transition from
  steady to weakly turbulent flow in a close-packed ordered array of spheres}.
  \jt{Journal of Fluid Mechanics}  \bvol{465},  \pg{59--97}.

\bibitem[Jin {\em et~al.\/}(2015)Jin, Uth, Kuznetsov \& Herwig]{jin2015}
{\sc \au{Jin, Y.}, \au{Uth, M.~F.}, \au{Kuznetsov, A.~V.} \& \au{Herwig, H.}}
  \yr{2015}  \at{Numerical investigation of the possibility of macroscopic
  turbulence in porous media: a direct numerical simulation study}.
  \jt{Journal of Fluid Mechanics}  \bvol{766},  \pg{76--103}.

\bibitem[Mahmud {\em et~al.\/}(2019)Mahmud, Van~Hong \&
  Lestariono]{mahmud2019sand}
{\sc \au{Mahmud, Hisham~Ben}, \au{Van~Hong, Leong} \& \au{Lestariono, Yuli}}
  \yr{2019}  \at{Sand production: A smart control framework for risk
  mitigation}.  \jt{Petroleum}  \bvol{6}~(1),  \pg{1--13}.

\bibitem[Mallat(2009)]{Mallat1999}
{\sc \au{Mallat, Stéphane}} \yr{2009} {\em A wavelet tour of signal
  processing\/}, 3rd edn.  \publ{Boston: Academic Press}.

\bibitem[Matsuda {\em et~al.\/}(2014)Matsuda, Onishi, Hirahara, Kurose,
  Takahashi \& Komori]{matsuda2014influence}
{\sc \au{Matsuda, Keigo}, \au{Onishi, Ryo}, \au{Hirahara, Masaaki}, \au{Kurose,
  Ryoichi}, \au{Takahashi, Keiko} \& \au{Komori, Satoru}} \yr{2014}
  \at{Influence of microscale turbulent droplet clustering on radar cloud
  observations}.  \jt{Journal of the Atmospheric Sciences}  \bvol{71}~(10),
  \pg{3569--3582}.

\bibitem[Matsuda {\em et~al.\/}(2021)Matsuda, Schneider \&
  Yoshimatsu]{matsuda2020scale}
{\sc \au{Matsuda, Keigo}, \au{Schneider, Kai} \& \au{Yoshimatsu, Katsunori}}
  \yr{2021}  \at{Scale-dependent statistics of inertial particle distribution
  in high reynolds number turbulence}.  \jt{Physical Review Fluids}  \bvol{6},
  \pg{064304}.

\bibitem[Maxey(1987)]{maxey1987gravitational}
{\sc \au{Maxey, Martin~R}} \yr{1987}  \at{The gravitational settling of aerosol
  particles in homogeneous turbulence and random flow fields}.  \jt{Journal of
  Fluid Mechanics}  \bvol{174},  \pg{441--465}.

\bibitem[Meneveau(1991)]{meneveau1991analysis}
{\sc \au{Meneveau, Charles}} \yr{1991}  \at{Analysis of turbulence in the
  orthonormal wavelet representation}.  \jt{Journal of Fluid Mechanics}
  \bvol{232},  \pg{469--520}.

\bibitem[Monchaux {\em et~al.\/}(2012)Monchaux, Bourgoin \&
  Cartellier]{monchaux2012analyzing}
{\sc \au{Monchaux, Romain}, \au{Bourgoin, Mickael} \& \au{Cartellier, Alain}}
  \yr{2012}  \at{Analyzing preferential concentration and clustering of
  inertial particles in turbulence}.  \jt{International Journal of Multiphase
  Flow}  \bvol{40},  \pg{1--18}.

\bibitem[Monin \& Yaglom(1965)]{monin1965}
{\sc \au{Monin, A.~S.} \& \au{Yaglom, A.~M.}} \yr{1965} {\em Statistical fluid
  mechanics: Mechanics of Turbulence\/}.  \publ{The MIT Press}.

\bibitem[Orlandi {\em et~al.\/}(2018)Orlandi, Davide \&
  Pirozzoli]{orlandi2018dns}
{\sc \au{Orlandi, Paolo}, \au{Davide, Modesti} \& \au{Pirozzoli, Sergio}}
  \yr{2018}  \at{{D}{N}{S} of turbulent flows in ducts with complex shape}.
  \jt{Flow, Turbulence and Combustion}  \bvol{100}~(4),  \pg{1063--1079}.

\bibitem[Oujia {\em et~al.\/}(2020)Oujia, Matsuda \& Schneider]{oujia2020jfm}
{\sc \au{Oujia, Thibault}, \au{Matsuda, Keigo} \& \au{Schneider, Kai}}
  \yr{2020}  \at{Divergence and convergence of inertial particles in high
  {R}eynolds number turbulence}.  \jt{Journal of Fluid Mechics}  \bvol{905},
  \pg{A14}.

\bibitem[Pandya {\em et~al.\/}(1998)Pandya, Bhuniya \&
  Khilar]{pandya1998existence}
{\sc \au{Pandya, Vishal~B}, \au{Bhuniya, S} \& \au{Khilar, Kartic~C}} \yr{1998}
   \at{Existence of a critical particle concentration in plugging of a packed
  bed}.  \jt{American Institute of Chemical Engineers. AIChE Journal}
  \bvol{44}~(4),  \pg{978}.

\bibitem[Patil \& Liburdy(2013)]{patil2013}
{\sc \au{Patil, V.~A.} \& \au{Liburdy, J.~A.}} \yr{2013}  \at{Flow structures
  and their contribution to turbulent dispersion in a randomly packed porous
  bed based on particle image velocimetry measurements}.  \jt{Physics of
  Fluids}  \bvol{25}~(11),  \pg{24}.

\bibitem[Ramachandran \& Fogler(1999)]{ramachandran1999plugging}
{\sc \au{Ramachandran, Venkatachalam} \& \au{Fogler, H~Scott}} \yr{1999}
  \at{Plugging by hydrodynamic bridging during flow of stable colloidal
  particles within cylindrical pores}.  \jt{Journal of Fluid Mechanics}
  \bvol{385},  \pg{129--156}.

\bibitem[Saucier(1974)]{saucier1974considerations}
{\sc \au{Saucier, RJ}} \yr{1974}  \at{Considerations in gravel pack design}.
  \jt{Journal of Petroleum Technology}  \bvol{26}~(02),  \pg{205--212}.

\bibitem[Schneider {\em et~al.\/}(2004)Schneider, Farge \&
  Kevlahan]{schneider2004spatial}
{\sc \au{Schneider, Kai}, \au{Farge, Marie} \& \au{Kevlahan, Nicholas}}
  \yr{2004}  \at{Spatial intermittency in two-dimensional turbulence: a wavelet
  approach}.  \bt{In {\em Woods Hole Mathematics: Perspectives in Mathematics
  and Physics\/}},  \pg{pp. 302--328}.  \publ{World Scientific}.

\bibitem[Schneider \& Vasilyev(2010)]{schneidervasilyev2010}
{\sc \au{Schneider, Kai} \& \au{Vasilyev, Oleg~V}} \yr{2010}  \at{Wavelet
  methods in computational fluid dynamics}.  \jt{Annual Review of Fluid
  Mechanics}  \bvol{42},  \pg{473--503}.

\bibitem[Shams {\em et~al.\/}(2013)Shams, Roelofs, Komen \&
  Baglietto]{shams2013quasi}
{\sc \au{Shams, A}, \au{Roelofs, F}, \au{Komen, EMJ} \& \au{Baglietto, E}}
  \yr{2013}  \at{Quasi-direct numerical simulation of a pebble bed
  configuration. part i: Flow (velocity) field analysis}.  \jt{Nuclear
  Engineering and Design}  \bvol{263},  \pg{473--489}.

\bibitem[Sundaram \& Collins(1997)]{sundaram1997collision}
{\sc \au{Sundaram, Shivshankar} \& \au{Collins, Lance~R}} \yr{1997}
  \at{Collision statistics in an isotropic particle-laden turbulent suspension.
  part 1. direct numerical simulations}.  \jt{Journal of Fluid Mechanics}
  \bvol{335},  \pg{75--109}.

\bibitem[Toschi \& Bodenschatz(2009)]{toschi2009lagrangian}
{\sc \au{Toschi, Federico} \& \au{Bodenschatz, Eberhard}} \yr{2009}
  \at{Lagrangian properties of particles in turbulence}.  \jt{Annual Review of
  Fluid Mechanics}  \bvol{41},  \pg{375--404}.

\bibitem[Valdes \& Santamarina(2006)]{valdes2006particle}
{\sc \au{Valdes, Julio~R} \& \au{Santamarina, J~Carlos}} \yr{2006}
  \at{Particle clogging in radial flow: Microscale mechanisms}.  \jt{SPE
  Journal}  \bvol{11}~(02),  \pg{193--198}.

\bibitem[Wang {\em et~al.\/}(2000)Wang, Wexler \& Zhou]{wang2000statistical}
{\sc \au{Wang, Lian-Ping}, \au{Wexler, Anthony~S} \& \au{Zhou, Yong}} \yr{2000}
   \at{Statistical mechanical description and modelling of turbulent collision
  of inertial particles}.  \jt{Journal of Fluid Mechanics}  \bvol{415},
  \pg{117--153}.

\bibitem[Wood {\em et~al.\/}(2020)Wood, He \& Apte]{wood2020modeling}
{\sc \au{Wood, Brian~D.}, \au{He, Xiaoliang} \& \au{Apte, Sourabh~V.}}
  \yr{2020}  \at{Modeling turbulent flows in porous media}.  \jt{Annual Review
  of Fluid Mechanics}  \bvol{52}~(1),  \pg{171--203},  \arxiv{arXiv:
  https://doi.org/10.1146/annurev-fluid-010719-060317}.

\bibitem[Yi {\em et~al.\/}(2005)Yi, Valk{\'o} \& Russell]{yi2005effect}
{\sc \au{Yi, Xi}, \au{Valk{\'o}, PP} \& \au{Russell, J.E.}} \yr{2005}
  \at{Effect of rock strength criterion on the predicted onset of sand
  production}.  \jt{International Journal of Geomechanics}  \bvol{5}~(1),
  \pg{66--73}.

\end{thebibliography}

\end{document}